\documentclass[pra,prl,twocolumn,showpacs,floatfix,10pt,twoside]{revtex4-1}
\usepackage{graphicx}
\usepackage{dcolumn}
\usepackage{bm}
\usepackage{subfigure}
\usepackage{float}
\usepackage{multirow}
\usepackage{booktabs}
\usepackage{amsmath}
\usepackage{lipsum}
\usepackage{latexsym,amssymb}
\usepackage[mathscr]{eucal}
\usepackage[switch*]{lineno}
\usepackage[bookmarks=true,
   colorlinks=true,
   linkcolor=blue,
   urlcolor=blue,
   citecolor=blue,
   bookmarks=true,
   hyperindex=true
]{hyperref}

\begin{document}
\allowdisplaybreaks[1]

\title{Joule-Thomson expansion of higher dimensional nonlinearly AdS black hole with power Maxwell invariant source}

\author{Zhong-Wen Feng\textsuperscript{1}}
\altaffiliation{Email: zwfengphy@163.com}
\author{Xia Zhou\textsuperscript{1}}
\author{Guansheng He\textsuperscript{3}}
\author{Shi-Qi Zhou\textsuperscript{1}}
\author{Shu-Zheng Yang\textsuperscript{1}}
\vskip 0.5cm
\affiliation{1 Physics and Space Science College, China West Normal University, Nanchong, 637009, China \\
2 School of Mathematics and Physics, University of South China, Hengyang, 421001, China}


\begin{abstract}
In this paper,  the Joule-Thomson expansion of the higher dimensional nonlinearly AdS black hole with power Maxwell invariant source is investigated. The results show the Joule-Thomson coef\/f\/icient has a zero point and a divergent point, which are coincide with the inversion temperature $T_i$  and the zero point of Hawking temperature, respectively. The inversion temperature increases monotonously with inversion pressure. For high-pressure region, the inversion temperature decreases with the dimensionality $D$ and the nonlinearity parameter $s$, whereas it increases with the charge $Q$. However,  $T_i$ for low-pressure region increase with  $D$ and $s$, while it decreases with $Q$. The ratio ${\eta _{\rm{BH}}}$ between the minimum of inversion temperature and the critical temperature does not depend on  $Q$, it recovers the higher dimensional Reissner-N\"{o}rdstrom AdS black hole case when  $s=1$. However, for $s>1$, it becomes smaller and smaller as  $D$ increase and approaches a constant when $D\rightarrow\infty$. Finally, we found that increase of mass $M$ and $s$, or reduce the charge $Q$ and $D$ can enhance the isenthalpic curve, and the ef\/fect of $s$ on the isenthalpic curve is much greater than other parameters.
\end{abstract}
\keywords{Higher dimensional nonlinearly AdS black hole; Power Maxwell invariant source; Joule-Thomson Expansion}
\maketitle
\section{Introduction}
\label{Int}
By introducing the quantum mechanism into general relativity, Hawking and Bekenstein demonstrated that black holes have temperature and entropy \cite{cha1,cha2,cha3}. After that, people have established the complete theories of black hole thermodynamics. According to those theories, the thermodynamic properties of a variety of complicated black holes were explored in the past few decades \cite{chb1,chb2,chb3,chb4,chb5,chb6,chb7,chb8}. Among all, the anti-de Sitter (AdS) black holes have attracted people's attention for their strange thermodynamic properties. Hawking and Page f\/irst pointed out the existence of a thermodynamic phase transition in Schwarzschild-AdS spacetime \cite{cha4}. This heuristic work shows the deeper-seated relation between conf\/inement and deconf\/inement phase transition of the gauge f\/ield in the AdS/CFT correspondence. Within this context, a series of works have been done to investigate the common properties between AdS spacetimes and the general thermodynamic system. In Refs. \cite{cha5,cha6}, Chamblin \emph{et al}. proposed that the phase structures of Reissner-N\"{o}rdstrom(R-N) AdS black hole are similar to that of the Van der Waals system. Furthermore, by interpreting the cosmological constant $\Lambda$  and mass $M$  as the thermodynamic pressure $P$ and chemical enthalpy $H$, receptively \cite{cha7,cha8,cha9}, the implications of black hole thermodynamics have been investigated in many contexts. In the extended phase space, Kubiz\v{n}\'{a}k and Mann calculated the $P-V$  critical behavior of R-N AdS black hole and demonstrate that they also coincide with those of the Van der Waals f\/luid\cite{cha10}. In Ref.~\cite{cha11}, Johnson showed the AdS black holes can be considered as holographic heat engines, which outputs work from  a pressure-volume space. Besides, Dolan found that the black hole with the non-positive cosmological constant has no adiabatic compressibility \cite{cha12,cha13}.

There is no doubt now that the studying of the black holes' Van der Waals behavior is so important since it connects the gravity with the ordinary thermodynamical system. Recently, the black hole thermodynamics  have been extended to the regime of Joule-Thomson expansion \cite{chc1,chc2}. In the Van der Waals system, the Joule-Thomson expansion occurs when the gas from the high-pressure zone through the porous plug into the low-pressure zone. Meanwhile, the enthalpy remains the same in this process and can be used to def\/ine the non-equilibrium states of expansion. Based on the viewpoints of Refs.~\cite{cha8,cha14,cha15}, \"{O}kc\"{u} and Aydmer investigated the Joule-Thomson expansion of  R-N AdS black hole for the f\/irst time. Their result shows that R-N AdS black hole has a similar Joule-Thomson expansion process with the Van der Waals f\/luid  \cite{cha16}. Subsequently, the Joule-Thomson expansion of Kerr AdS black hole, $D$-dimensional R-N AdS spacetime, charged Gauss-Bonnet black hole, Bardeen-AdS black hole,  Hayward-AdS \emph{et al}. were studied in Refs.~\cite{cha17,cha18,cha19,cha20,cha21,cha22,cha22+,cha23+,cha23,cha24+,cha24++}.

By analyzing previous works, one  f\/inds that the in-depth analysis of Joule-Thomson expansion of higher-dimensional black holes in the complex gravity is still missing. To our knowledge, the reasons for this lie in two main aspects, one is  the Joule-Thomson expansion behavior of AdS black holes are closely related to their properties of spacetimes. The change of any parameter of spacetimes would make the calculations more complicated, and even unable to obtain accurate results. Moreover, the Van der Waals behavior is not valid for every AdS spacetimes \cite{chh1,chh2}, which may leads to the absence of Joule-Thomson expansion. Despite the dif\/f\/iculties, it benef\/icial to investigate how the dimension of spacetimes and the complex gravity af\/fect the Joule-Thomson expansion behavior since those works help people more understand the properties of black holes.

In recent years, an interest in the subject of the higher dimensional nonlinearly black hole has been growing. It is well known that, on the one hand, the black holes in the higher dimension have more physical information, which can be  detected by the Large Hadron Collider and Advanced LIGO's detection of gravitational waves \cite{chh3,chh4,chh5,chh6}. Besides, many unif\/ied theories show that the coupling characteristics between the gravity f\/ield and the gauge f\/ield can be obtained in higher dimensions. Hence, it is believed that the related researches open a new window into gravity and quantum gravity. On the other hand, many charge AdS black holes are derived from the Einstein-Maxwell-AdS theory. However, the classical Maxwell theory has some defects, such as the central singularity of the point-like charges, vacuum polarization in quantum electrodynamics \emph{et al.}, those leads to the linear electrodynamics fail when the electromagnetic f\/ield is too strong. To crackdown on those problems, Born and Infeld introduced the non-linear electrodynamics (NLEDs) by eliminating the inf\/inite self-energy. Furthermore, for generalizing NLEDs into higher dimensions and keeping the conformal symmetry of Maxwell action, one proposed the power Maxwell invariant (PMI) theory, which is an improved NLEDs model \cite{cha24,cha25}. Therefore, the Lagrangian density of PMI f\/ield  can be expressed as  ${L_{{\rm{PMI}}}} = {\left( { - F} \right)^s} = {\left( { - {F_{\mu \nu }}{F^{\mu \nu }}} \right)^s}$ with an arbitrary rational number  $s$.  It causes the properties of  PMI f\/ield are more abundant than the Maxwell f\/ield. By introducing the PMI source into the astrophysical frameworks, many new results are obtained \cite{cha26,cha27,cha28,cha28+}. Moreover, the source turns out to be important in the studying on the strongly coupled dual gauge theory \cite{chh7}. Recently, the black hole solutions Einstein gravity coupled to the PMI theory have attracted people's attention since they believe the non-linear properties of the fundamental theory can be found in those spacetimes. In Refs.~\cite{cha29,cha31,cha32}, the thermodynamic properties of those black holes have been analyzed in detail. In particular, in the extended phase space, the phase structure and the critical behavior of higher dimensional nonlinearly AdS black hole with PMI source are similar to those of Van der Waals f\/luid \cite{cha33}. Due to the above discussion, it is believed that the higher dimensional nonlinearly AdS black hole  with PMI source has the Joule-Thomson expansion process. Therefore, we calculate Joule-Thomson coef\/f\/icient, the equation of the inversion curve and the equation of the isenthalpic curves in this paper. Meanwhile, we also use the numerical method to analyze the inf\/luence of dimensionality $D$, and the nonlinearity parameter $s$ on the Joule-Thomson expansion.

The rest of the paper is organized as follows. In the next section, we review the metric of nonlinearly AdS black hole and its thermodynamic properties. Section~3 is devoted to deriving the Joule-Thomson coef\/f\/icient, the ratio between the minimum of inversion temperature and the critical temperature, the equation of the inversion curve, and the equation of the isenthalpic curve. Then, we investigate the influence of various parameters (especially the dimensionality and the nonlinearity parameter) of the black hole on Joule-Thomson expansion via the numerical method. The conclusion and discussion are contained in Section~4. This research takes the units  $G = c = {k_B} = 1$.

\section{The higher dimensional nonlinearly charged black hole with PMI source}
\label{sce2}
\subsection{The metric of higher dimensional nonlinearly AdS black hole with PMI source}
To begin with, it is necessary to review the metric and the thermodynamic properties of nonlinearly AdS black hole with PMI source. In the $D$-dimensional ($D\geq4$) AdS spacetimes, the bulk action of Einstein-PMI gravity can be written in the form \cite{cha24,cha25}:
\begin{align}
 \label{eq1}
I =  - \frac{1}{{16\pi }}\int {{{\rm{d}}^D}} x\sqrt { - g} \left[ {R - 2\Lambda  + {{\left( { - {F_{\mu \nu }}{F^{\mu \nu }}} \right)}^s}} \right],
\end{align}
where $R$ is the Ricci scalar,  $\Lambda  = {{ - \left( {D - 1} \right)\left( {D - 2} \right)} \mathord{\left/ {\vphantom {{ - \left( {D - 1} \right)\left( {D - 2} \right)} {2{l^2}}}} \right.
 \kern-\nulldelimiterspace} {2{l^2}}}$ is the cosmological constant with the radius of the AdS space $l$,  ${F_{\mu \nu }} = {\partial _\mu }{A_\nu } - {\partial _\nu }{A_\mu }$  represents the strength of the electromagnetic f\/ield, and $s$  is the nonlinearity parameter. Based on the Eq.~(\ref{eq1}), the line element of a $D$-dimensional nonlinearly AdS black hole with PMI source takes the form \cite{cha33}
\begin{align}
 \label{eq2}
{\rm{d}}{s^2} =  - f\left( r \right){\rm{d}}{t^2} + \frac{{{\rm{d}}{r^2}}}{{f\left( r \right)}} + {r^2}{\rm{d}}\Omega _{D - 2}^2,
\end{align}
where ${\rm{d}}\Omega _{D - 2}^2$  denotes the line element of a $\left(D-2\right)$-dimensional space. When considering the f\/ield equations arising from the variation of the bulk action with the metric~(\ref{eq2}), the metric function  $f\left( r \right)$ and gauge potential $A$  are given by:
\begin{align}
 \label{eq3}
f\left( r \right) = 1 - \frac{m}{{{r^{D - 3}}}} + \frac{{{r^2}}}{{{l^2}}} - \frac{{{{\left( {2s - 1} \right)}^2}{{\left[ {\frac{{\left( {D - 2} \right){{\left( {2s - D + 1} \right)}^2}{q^2}}}{{\left( {D - 3} \right){{\left( {2s - 1} \right)}^2}}}} \right]}^s}}}{{\left( {D - 2} \right)\left( {2s - D + 1} \right){r^{2\left( {\frac{{Ds - 4s + 1}}{{2s - 1}}} \right)}}}},
\end{align}
\begin{align}
 \label{eq4}
A =  - \sqrt {\frac{{D - 2}}{{2\left( {D - 3} \right)}}} q{r^{\frac{{2s - D + 1}}{{2s - 1}}}}{\rm{d}}t,
\end{align}
where the electromagnetic f\/ield 1-form is  $F = {\rm{d}}A$. Notably, to keep the nonlinear term of the source, the nonlinear parameter should satisfy  $s > {1 \mathord{\left/ {\vphantom {1 2}} \right.  \kern-\nulldelimiterspace} 2}$  and  $s \ne {{\left( {D - 1} \right)} \mathord{\left/ {\vphantom {{\left( {D - 1} \right)} 2}} \right. \kern-\nulldelimiterspace} 2}$. The parameters  $m$ and $q$  are related to the ADM mass $M$ and total electric charge $Q$ of the black hole respectively, which reads
\begin{align}
 \label{eq5}
M = & \frac{{{\omega _{D - 2}}\left( {D - 2} \right)}}{{16\pi }}m,
 \nonumber \\
Q = & \frac{{{\omega _{D - 2}}\sqrt 2 (2s - 1)s}}{{8\pi }}{\left( {\frac{{D - 2}}{{D - 3}}} \right)^{s - \frac{1}{2}}}{\left[ {\frac{{\left( {D - 1 - 2s} \right)q}}{{2s - 1}}} \right]^{2s - 1}},
\end{align}
with the volume of a unit $\left(D-2\right)$ sphere  ${\omega _{D - 2}} = {{2{\pi ^{\frac{{D - 1}}{2}}}} \mathord{\left/ {\vphantom {{2{\pi ^{\frac{{D - 1}}{2}}}} {\Gamma \left[ {{{\left( {D - 1} \right)} \mathord{\left/ {\vphantom {{\left( {D - 1} \right)} 2}} \right. \kern-\nulldelimiterspace} 2}} \right]}}} \right. \kern-\nulldelimiterspace} {\Gamma \left[ {{{\left( {D - 1} \right)} \mathord{\left/ {\vphantom {{\left( {D - 1} \right)} 2}} \right. \kern-\nulldelimiterspace} 2}} \right]}}$. According to  Eq.~(\ref{eq5}), metric function~(\ref{eq3}) can be rewrit\/ten as follows:
\begin{align}
 \label{eq6}
 f\left( r \right)  = &  1 - \frac{{16M\pi }}{{\left( {D - 2} \right){r^{D - 3}}{\omega _{D - 2}}}} + \frac{{16P\pi {r^2}}}{{\left( {D - 1} \right)\left( {D - 2} \right)}}  
  \nonumber \\
 & + \frac{{{{\left( {1 - 2s} \right)}^2}}}{{\left( {D - 2} \right)\left( {1 - D + 2s} \right)}}{r^{\left[ {\frac{{2 + 2s\left( {D - 4} \right)}}{{1 - 2s}}} \right]}}\Theta ,
\end{align}
where
\begin{align}
 \label{eq7}
\Theta  = {\left\{ {\frac{{{2^{\frac{5}{{2s - 1}}}}{\pi ^{\frac{2}{{2s - 1}}}}\left( {D - 2} \right)}}{{D - 3}}{{\left[ {\frac{{{{\left( {1 + \frac{1}{{D - 3}}} \right)}^{\frac{1}{2} - s}}Q}}{{s\left( {2s - 1} \right){\omega _{D - 2}}}}} \right]}^{\frac{2}{{2s - 1}}}}} \right\}^s},
\end{align}
and  $P = {{ - \Lambda } \mathord{\left/ {\vphantom {{ - \Lambda } {8\pi }}} \right. \kern-\nulldelimiterspace} {8\pi }} = {{(D - 1)(D - 2)} \mathord{\left/ {\vphantom {{(D - 1)(D - 2)} {16\pi {l^2}}}} \right. \kern-\nulldelimiterspace} {16\pi {l^2}}}$ is a thermodynamic pressure, which is a key def\/inition in the framework of black hole chemistry.

\subsection{The thermodynamics of higher dimensional nonlinearly charged AdS black hole with PMI source}
On the event horizon  $r_+$, one can rewrite the ADM mass in terms of the $P$  with the condition ${\left. {f\left( r \right)} \right|_{r = r}}_{_ + } = 0$, which reads as follows:
\begin{align}
 \label{eq8}
M   = &  - \frac{{r_ + ^{D - 3}{\omega _{D - 2}}}}{{8\pi }}\left[ {1 - \frac{D}{2} - \frac{{8P\pi r_ + ^2}}{{D - 1}}} \right. 
  \nonumber \\
& \left. { + \frac{{{{\left( {1 - 2s} \right)}^2}\Theta }}{{2\left( {1 - D + 2s} \right)}}r_ + ^{\frac{{2 + 2s\left( {D - 4} \right)}}{{1 - 2s}}}} \right].
\end{align}
The Hawking temperature of the black hole can be easily obtained as
\begin{align}
 \label{eq9}
T = & \frac{{f'\left( {{r_ + }} \right)}}{{4\pi }} 
  \nonumber \\
= & \frac{1}{{4\pi }}\left[ {\frac{{D - 3}}{r} + \frac{{16P\pi r}}{{D - 2}}} \right. \left. { - \frac{{\left( {2s - 1} \right)\Theta }}{{D - 2}}{r^{\frac{{1 + 2s\left( {D - 3} \right)}}{{1 - 2s}}}}} \right],
\end{align}
and the entropy is
\begin{align}
 \label{eq10}
S = {\int_0^{{r_ + }} {\frac{1}{T}\left( {\frac{{\partial M}}{{\partial r}}} \right)} _{Q,P}}dr = \frac{{{\omega _{D - 2}}}}{4}r_ + ^{D - 2}.
\end{align}
It is clear that the entropy of the black hole obeys the area formula $S = {\mathcal{A} \mathord{\left/ {\vphantom {\mathcal{A} 4}} \right. \kern-\nulldelimiterspace} 4}$ with the area of the black hole $\mathcal{A}$. Based on the above thermodynamics quantities, the f\/irst law of black hole thermodynamics in the extended phase space is given by
\begin{align}
 \label{eq11}
{\rm{d}}M = T{\rm{d}}S + V{\rm{d}}P + \Phi {\rm{d}}Q,
\end{align}
where  $\Phi  ={\left( {{{\partial M} \mathord{\left/ {\vphantom {{\partial M} {\partial Q}}} \right. \kern-\nulldelimiterspace} {\partial Q}}} \right)_{S,P}}= \frac{{{2^{\frac{{3 - s}}{{2s - 1}}}}{\pi ^{\frac{1}{{2s - 1}}}}\left( {1 - 2s} \right)}}{{1 - D{\rm{ + }}2s}}$ ${\Theta ^{\frac{1}{s}}} r_ + ^{\frac{{1 - D + 2s}}{{2D - 1}}}\sqrt {\frac{{D - 2}}{{D - 3}}} $ is the electric potential, and  $V = {\left( {{{\partial M} \mathord{\left/ {\vphantom {{\partial M} {\partial P}}} \right. \kern-\nulldelimiterspace} {\partial P}}} \right)_{S,P}} = {{{\omega _{D - 2}}r_ + ^{D - 1}} \mathord{\left/ {\vphantom {{{\omega _{D - 2}}r_ + ^{D - 1}} {\left( {D - 1} \right)}}} \right. \kern-\nulldelimiterspace} {\left( {D - 1} \right)}}$ is the thermodynamic volume, which corresponding conjugate quantity is the thermodynamic pressure  $P$ \cite{cha34}. The connected Smarr relation becomes
\begin{align}
 \label{eq12}
\left( {D - 3} \right)M = \left( {D - 2} \right)TS - 2PV + \left( {D - 3} \right)\Phi Q.
\end{align}
Next, substituting Eq.~(\ref{eq8}) into Eq.~(\ref{eq9}),  the equation of state is
\begin{align}
 \label{eq13}
P =  - \frac{{\left( {D - 2} \right)\left( {D - 3} \right)}}{{16\pi r_ + ^2}} + \frac{{\left( {D - 2} \right)T}}{{4{r_ + }}} + \frac{{\left( {2s - 1} \right)\Theta }}{{16\pi }}r_ + ^{ - \frac{{2s\left( {D - 2} \right)}}{{2s - 1}}} .
\end{align}
According to the viewpoints in Ref.~\cite{cha10}, the critical points of this thermodynamic system can be obtained by the conditions ${\left( {{{\partial P} \mathord{\left/ {\vphantom {{\partial P} {\partial {r_ + }}}} \right. \kern-\nulldelimiterspace} {\partial {r_ + }}}} \right)_{T = {T_{cr}}}} = {\left( {{{{\partial ^2}P} \mathord{\left/ {\vphantom {{{\partial ^2}P} {\partial r_ + ^2}}} \right. \kern-\nulldelimiterspace} {\partial r_ + ^2}}} \right)_{T = {T_{cr}}}} = 0$, and the critical temperature is given by
\begin{align}
 \label{eq14}
{T_{cr}} = & \frac{{4\left( {D - 3} \right)\left( {Ds - 4s + 1} \right)}}{{\pi \left( {D - 2} \right)\left( {2Ds - 6s + 1} \right)}}
   \nonumber \\
 & \times {\left[ {\frac{{ks{{\left( {D - 2} \right)}^2}\left( {2Ds - 6s + 1} \right){q^{2s}}}}{{16\left( {D - 3} \right){{\left( {2s - 1} \right)}^2}}}} \right]^{\frac{{1 - 2s}}{{2(Ds - 4s + 1)}}}},
\end{align}
where
\begin{align}
 \label{eq15}
k = \frac{{{{16}^{\frac{{s\left( {D - 2} \right)}}{{2s - 1}}}}\left( {2s - 1} \right){{\left[ {\frac{{\left( {D - 2} \right){{\left( {2s - D + 1} \right)}^2}}}{{\left( {D - 3} \right){{\left( {2s - 1} \right)}^2}}}} \right]}^s}}}{{{{\left( {D - 2} \right)}^{\frac{{2s\left( {D - 2} \right)}}{{2s - 1}}}}}}.
\end{align}
Here we only express the critical temperature since it will be used to analyze the Joule-Thomson expansion of the black hole in the next section. The expressions critical pressure and the critical radius can be found in Ref.~\cite{cha33} if needed. From Eq.~(\ref{eq9})-Eq.~(\ref{eq15}), one can see that the thermodynamic quantities of the higher dimensional nonlinearly AdS black hole with PMI source sensitively depend on the event horizon radius  $r_+$, the dimensionality  $D$, the ADM mass $M$, the total electric charge $Q$ and the nonlinearity parameter $s$. Remarkably, those thermodynamic quantities can easily go back to the higher dimensional R-N AdS case when $s=1$.

\section{Joule-Thomson expansion of higher dimensional nonlinearly AdS black hole with PMI source}
\label{sce3}
\begin{figure*}[htbp]
\centering
\subfigure[$Q=1, s=1.1$]{
\begin{minipage}[b]{0.3\textwidth}
\includegraphics[width=1.2\textwidth]{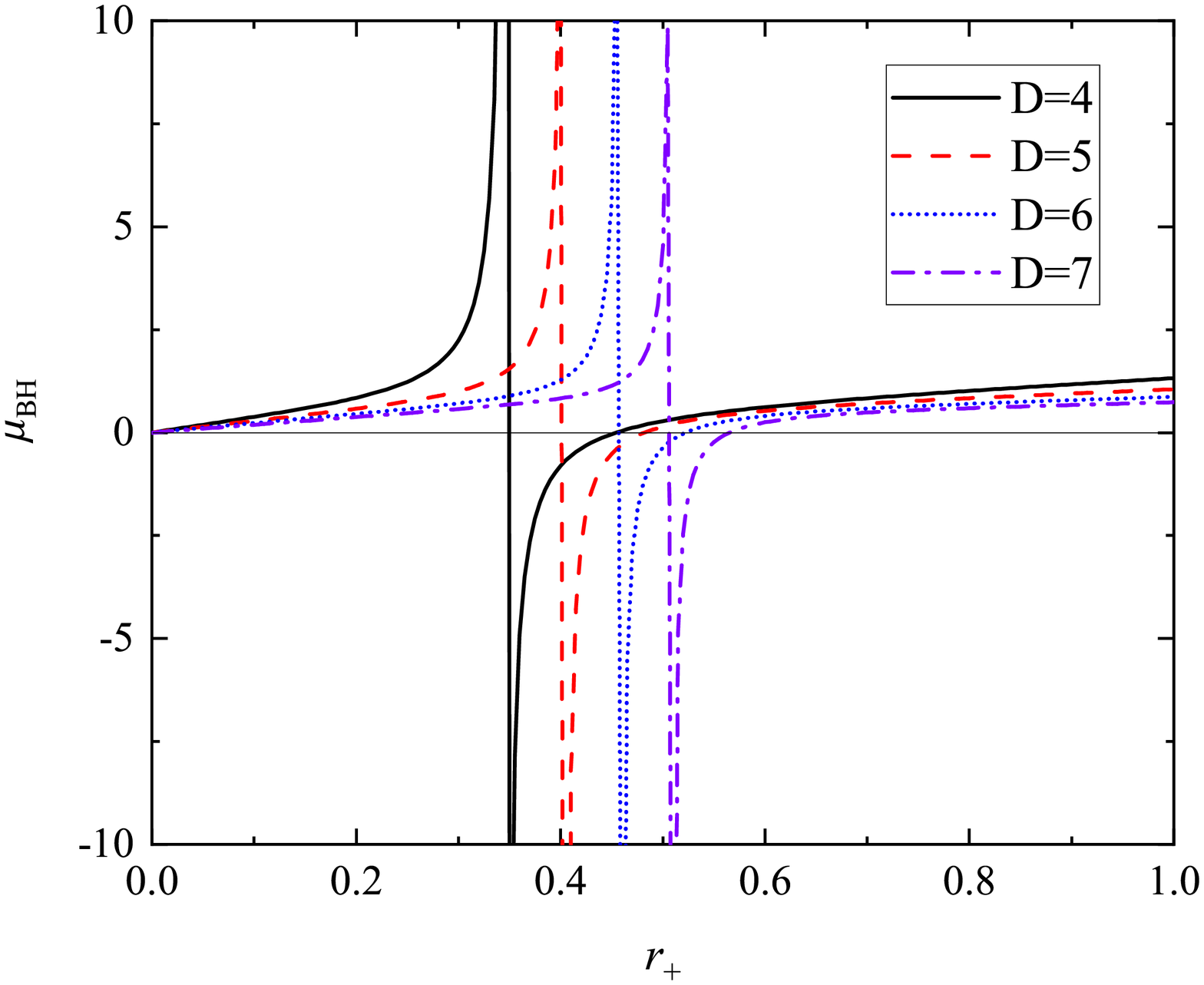}
\label{fig1-a}
\end{minipage}
}
\subfigure[$D=5,s=1.1$]{
\begin{minipage}[b]{0.3\textwidth}
\includegraphics[width=1.2\textwidth]{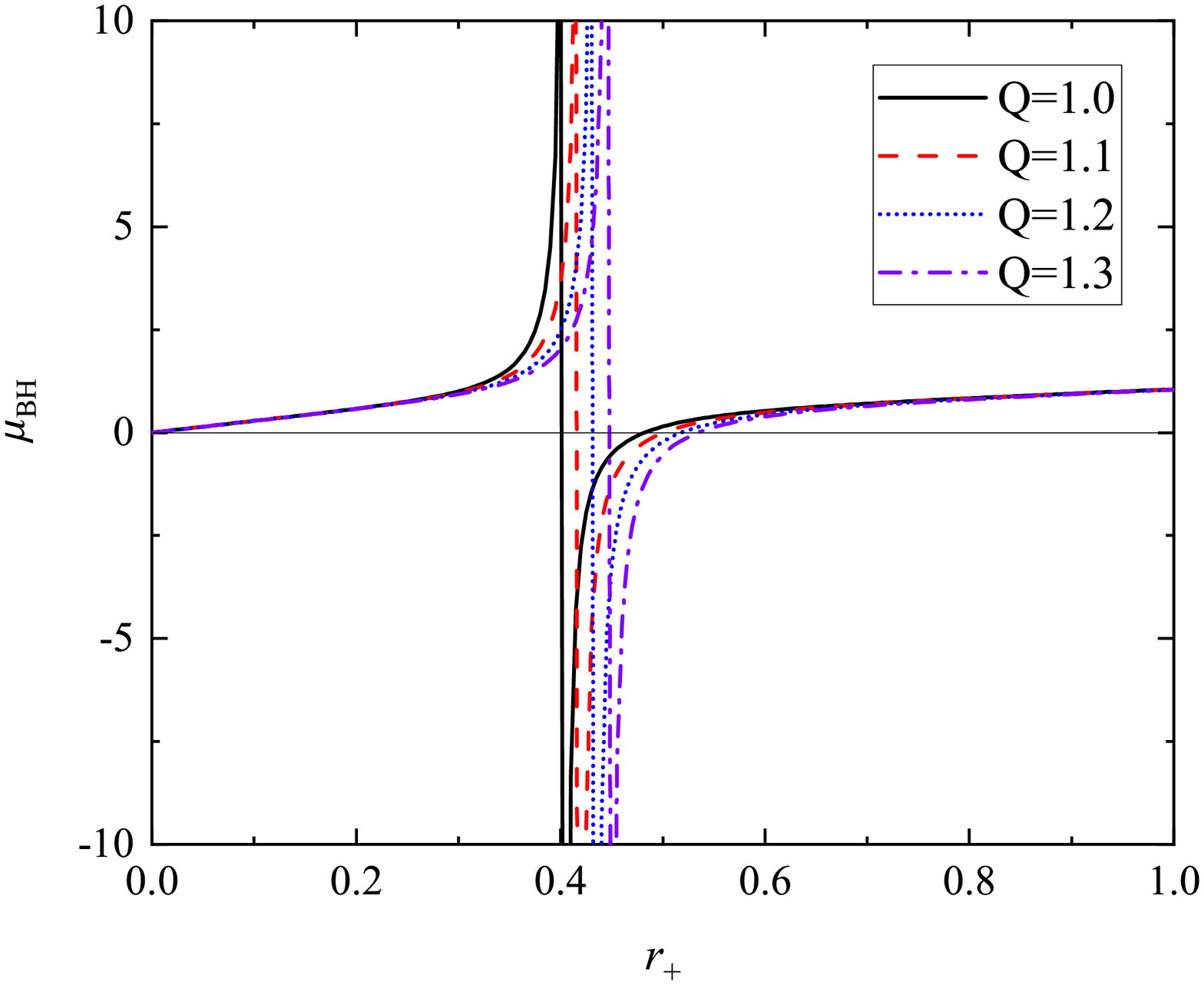}
\label{fig1-b}
\end{minipage}
}
\subfigure[$D=5,Q=1$]{
\begin{minipage}[b]{0.3\textwidth}
\includegraphics[width=1.2\textwidth]{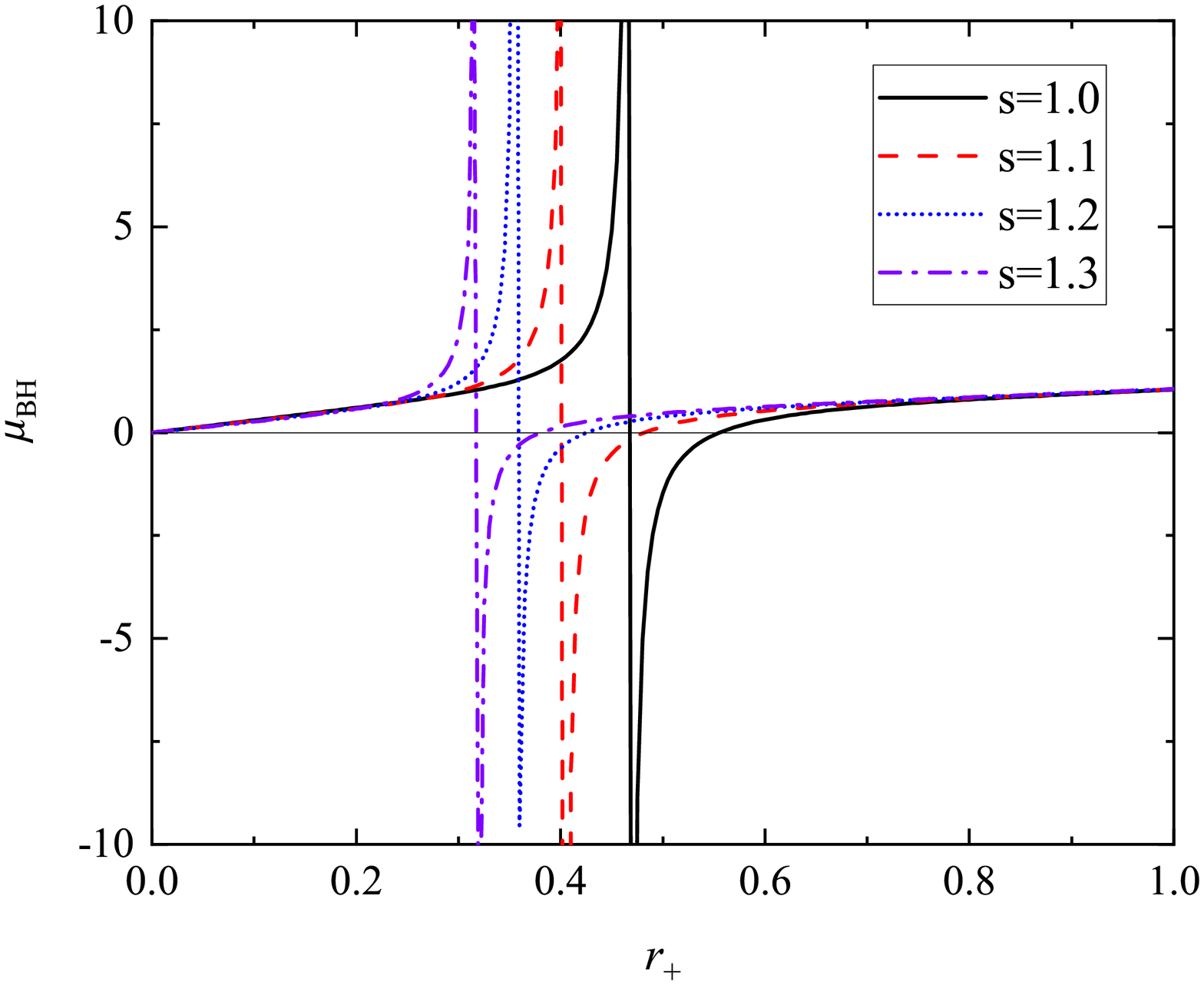}
\label{fig1-c}
\end{minipage}
}
\subfigure[$Q=1, s=1.1$]{
\begin{minipage}[b]{0.3\textwidth}
\includegraphics[width=1.2\textwidth]{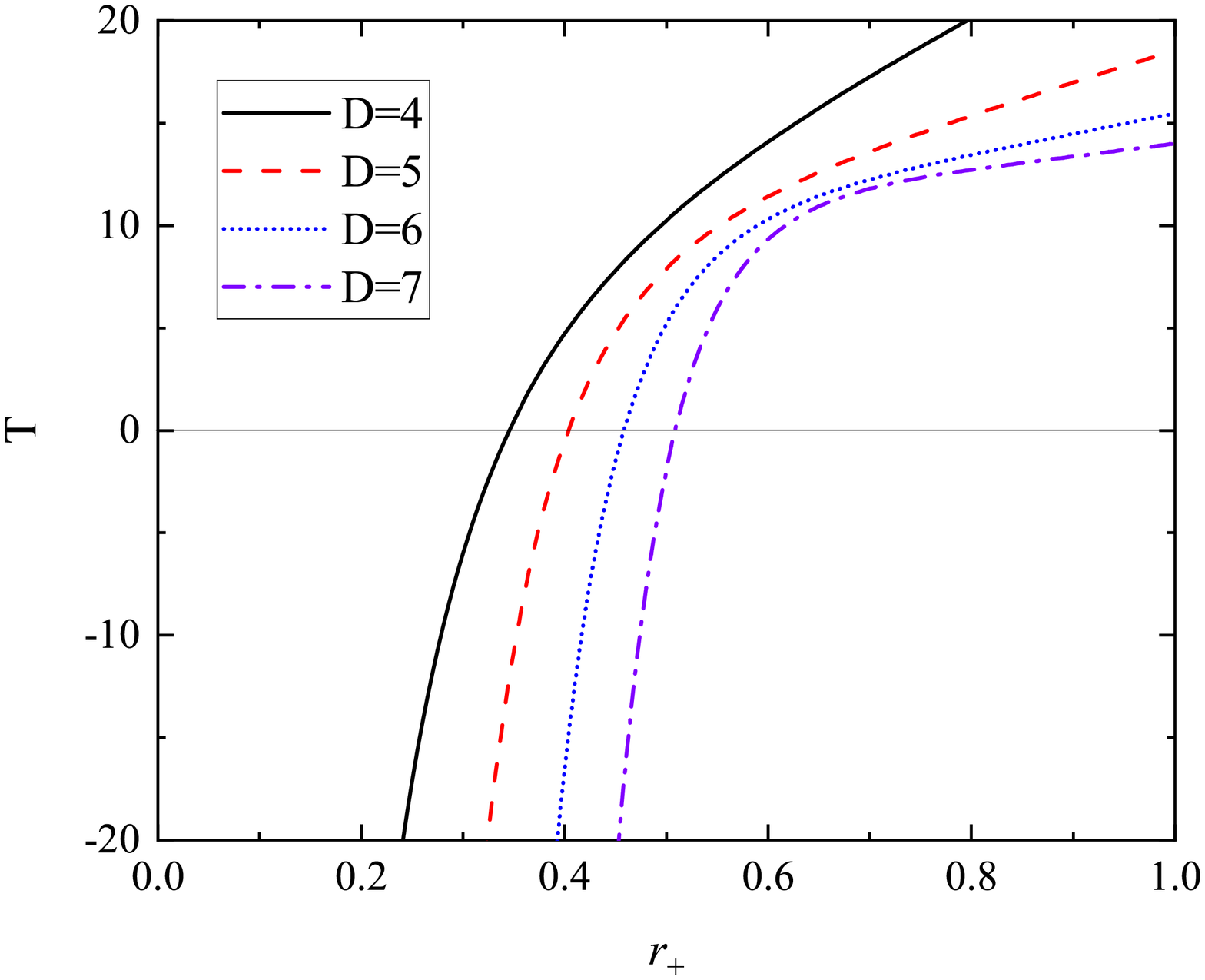}
\label{fig1-d}
\end{minipage}
}
\subfigure[$D=5, s=1.1$]{
\begin{minipage}[b]{0.3\textwidth}
\includegraphics[width=1.2\textwidth]{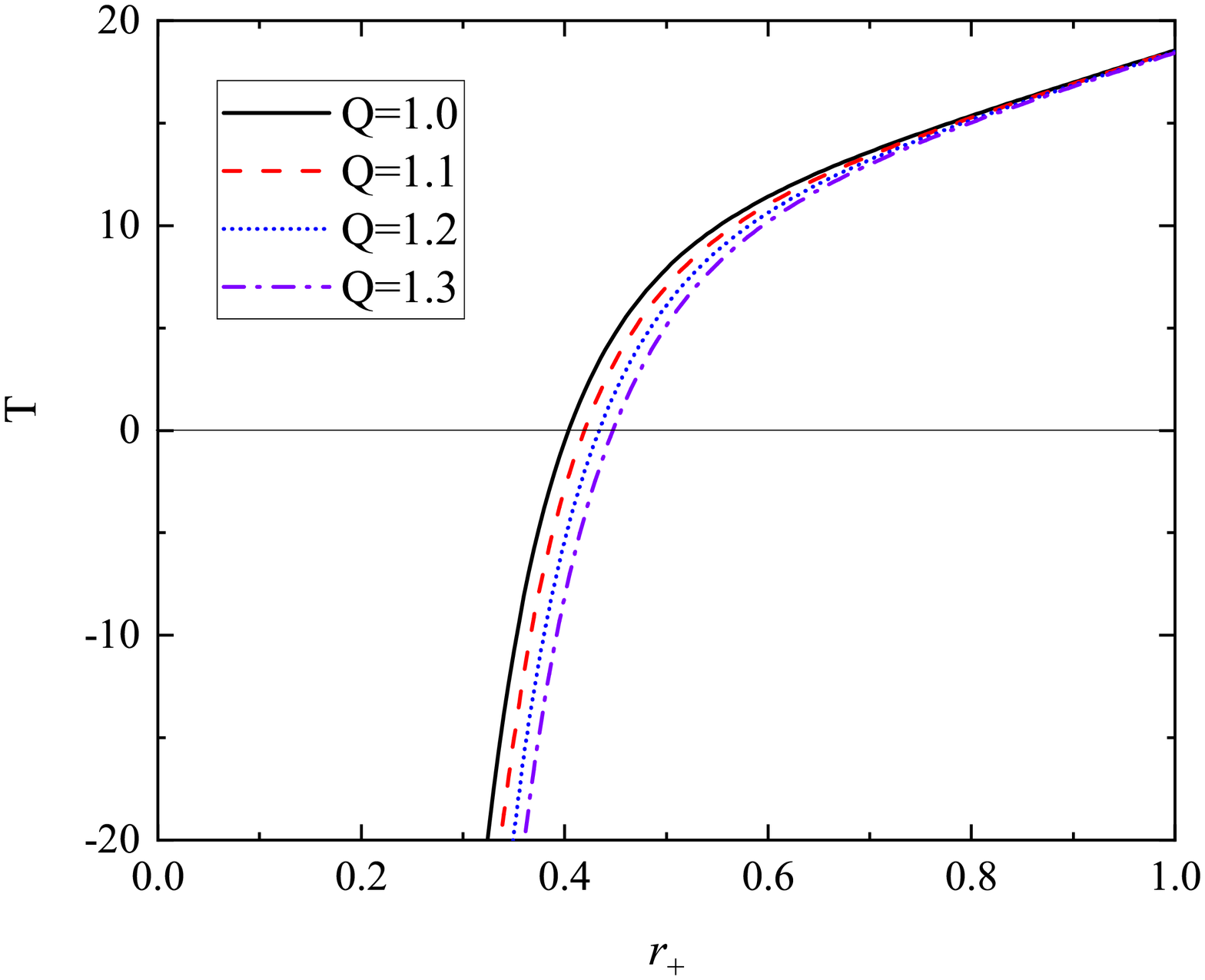}
\label{fig1-e}
\end{minipage}
}
\subfigure[$D=5, Q=1$]{
\begin{minipage}[b]{0.3\textwidth}
\includegraphics[width=1.2\textwidth]{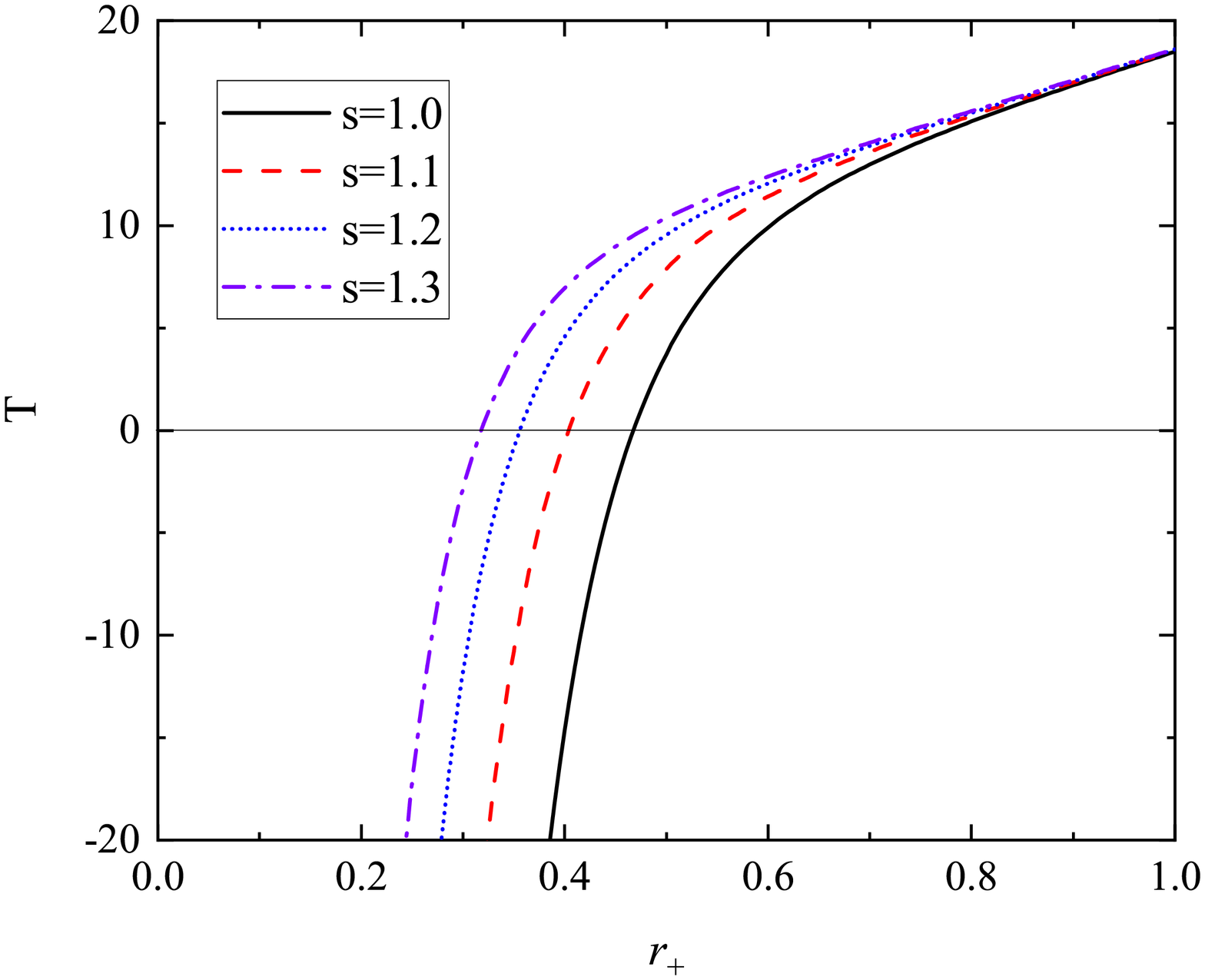}
\label{fig1-f}
\end{minipage}
}
\caption{(a)-(c) The inf\/luence of  $D$, $Q$  and  $s$ on ${\mu _{{\rm{BH}}}}$, respectively.  (d)-(f) The inf\/luence of  $D$, $Q$  and  $s$ on $T$, respectively.}
\label{fig1}
\end{figure*}

In this section, the Joule-Thomson expansion of higher dimensional nonlinearly AdS black hole with PMI source in the extended phase space is investigated. In the Joule-Thomson expansion, the enthalpy  $H$ of the Van der Waals system can be used to def\/ine the non-equilibrium states. Meanwhile, the enthalpy is related to the temperature and pressure of the system \cite{chc1,chc2}. Hence, one can f\/ind a slope of an isenthlpic curve in the $T - P$ plane, that is, the Joule-Thomson coef\/f\/icient. According to Ref.~\cite{cha16}, the Joule-Thomson coef\/f\/icient is denoted as follows:
\begin{align}
 \label{eq16}
\mu  = {\left( {\frac{{\partial T}}{{\partial P}}} \right)_H} = \frac{1}{{{C_p}}}\left[ {T{{\left( {\frac{{\partial V}}{{\partial T}}} \right)}_P} - V} \right],
\end{align}
where ${C_p} = T{\left( {{{\partial S} \mathord{\left/ {\vphantom {{\partial S} {\partial T}}} \right. \kern-\nulldelimiterspace} {\partial T}}} \right)_P}$  is the heat capacity at constant pressure. For  $\mu>0$, one has a cooling region in the $T - P$ plane, whereas a heating region appears for $\mu<0$. Moreover, when the Joule-Thomson coef\/f\/icient vanishes, one can obtain the inversion temperature ${T_i} = V{\left( {{{\partial T} \mathord{\left/ {\vphantom {{\partial T} {\partial V}}} \right. \kern-\nulldelimiterspace} {\partial V}}} \right)_P}$. In Ref.~\cite{cha33}, the phase structure of higher dimensional nonlinearly AdS black hole with PMI source is analogous to that of Van der Waals system. Thus, it is interesting to investigate the throttling process of the higher dimensional nonlinearly AdS black hole with PMI source. Now, substituting the thermodynamic quantities of the black hole into Eq.~(\ref{eq16}), one yields
\begin{widetext}
\begin{align}
 \label{eq17}
{\mu _{{\rm{BH}}}} = \frac{{4r_+\left[ {16P\pi r_ + ^2 + \left( {D - 3} \right)D - \Theta \left( {4s - 1} \right)r_ + ^{\frac{{2 + 2s\left( {D - 4} \right)}}{{1 - 2s}}}} \right]}}{{\left( {D - 1} \right)\left[ {6 + \left( {D - 5} \right)D + 16P\pi r_ + ^2 - \Theta \left( {1 - 2s} \right)r_ + ^{\frac{{2 + 2s\left( {D - 4} \right)}}{{1 - 2s}}}} \right]}}.
\end{align}
\end{widetext}
It is obvious that the Joule-Thomson coef\/f\/icient~(\ref{eq17}) is sensitive to the properties of spacetime of the black hole. By f\/ixing the pressure  $P$ and the charge $Q$, one can see how the dimensionality  $D$ and the nonlinearity parameter  $s$ af\/fect the behaviors of the Joule-Thomson coef\/f\/icient  ${\mu _{{\rm{BH}}}}$ and Hawking temperature $T$ in Fig.~\ref{fig1}.

From Fig.~\ref{fig1-a}-Fig.~\ref{fig1-c}, one can see that the general behavior of ${\mu _{{\rm{BH}}}}$ changes with  ${r_ + }$ as follows: when the event horizon enough large, ${\mu _{{\rm{BH}}}}$  is larger than zero. However, by decreasing the event horizon of the black hole, the Joule-Thomson coef\/f\/icient gradually decreases to zero (i. e.  $T_i$), and then goes to negative. Finally, the ${\mu _{{\rm{BH}}}}$ changes its sign at the divergent point. By comparing Figs.~\ref{fig1-a}-\ref{fig1-c} with Figs.~\ref{fig1-d}-\ref{fig1-f}, it is found that the Joule-Thomson coef\/f\/icient diverges at the points where the Hawking temperature becomes zero, and the ${\mu _{{\rm{BH}}}}$  decreases to zero as  ${r_ + } \to 0$. On the other hand, the curves of the Joule-Thomson coef\/f\/icient move to the right as the $D$ and $Q$ increase, while those curves move to the left as $s$  increases.

\begin{figure*}[htbp]
\centering
\subfigure[$D=4,s=1.1$]{
\begin{minipage}[b]{0.31\textwidth}
\includegraphics[width=1.25\textwidth]{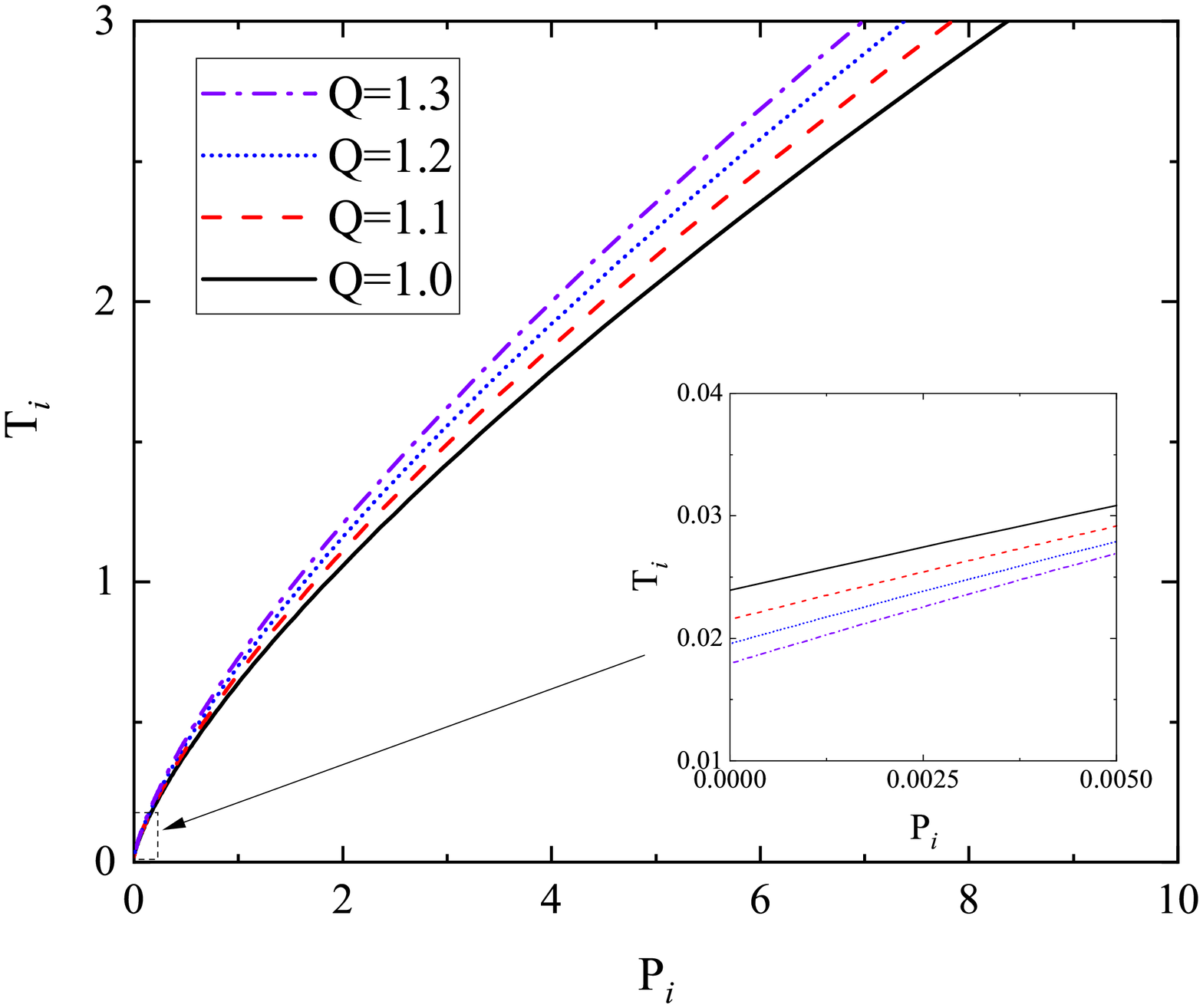}
\label{fig2-a}
\end{minipage}
}
\subfigure[$D=5,s=1.1$]{
\begin{minipage}[b]{0.31\textwidth}
\includegraphics[width=1.25\textwidth]{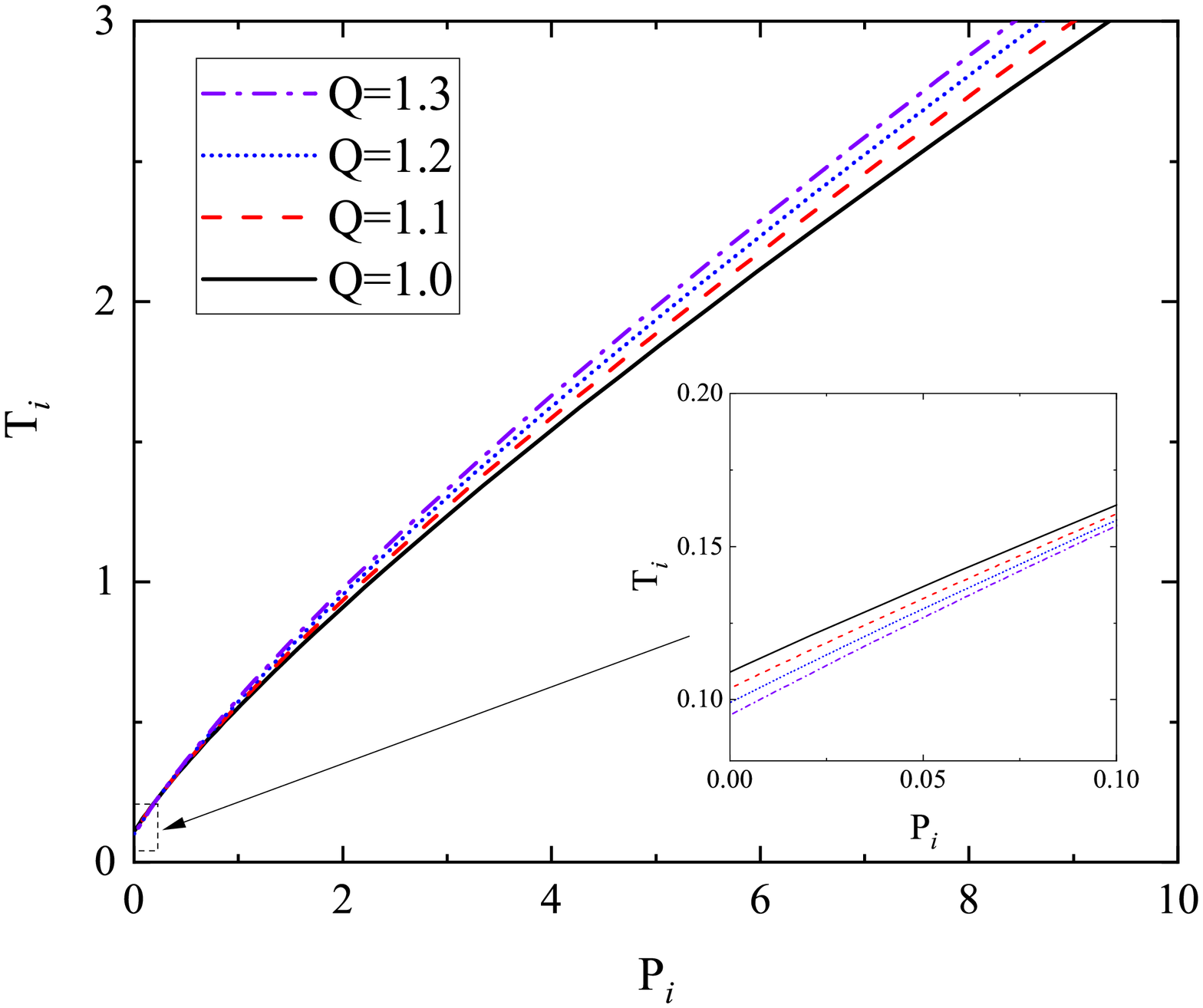}
\label{fig2-b}
\end{minipage}
}
\subfigure[$D=6,s=1.1$]{
\begin{minipage}[b]{0.31\textwidth}
\includegraphics[width=1.25\textwidth]{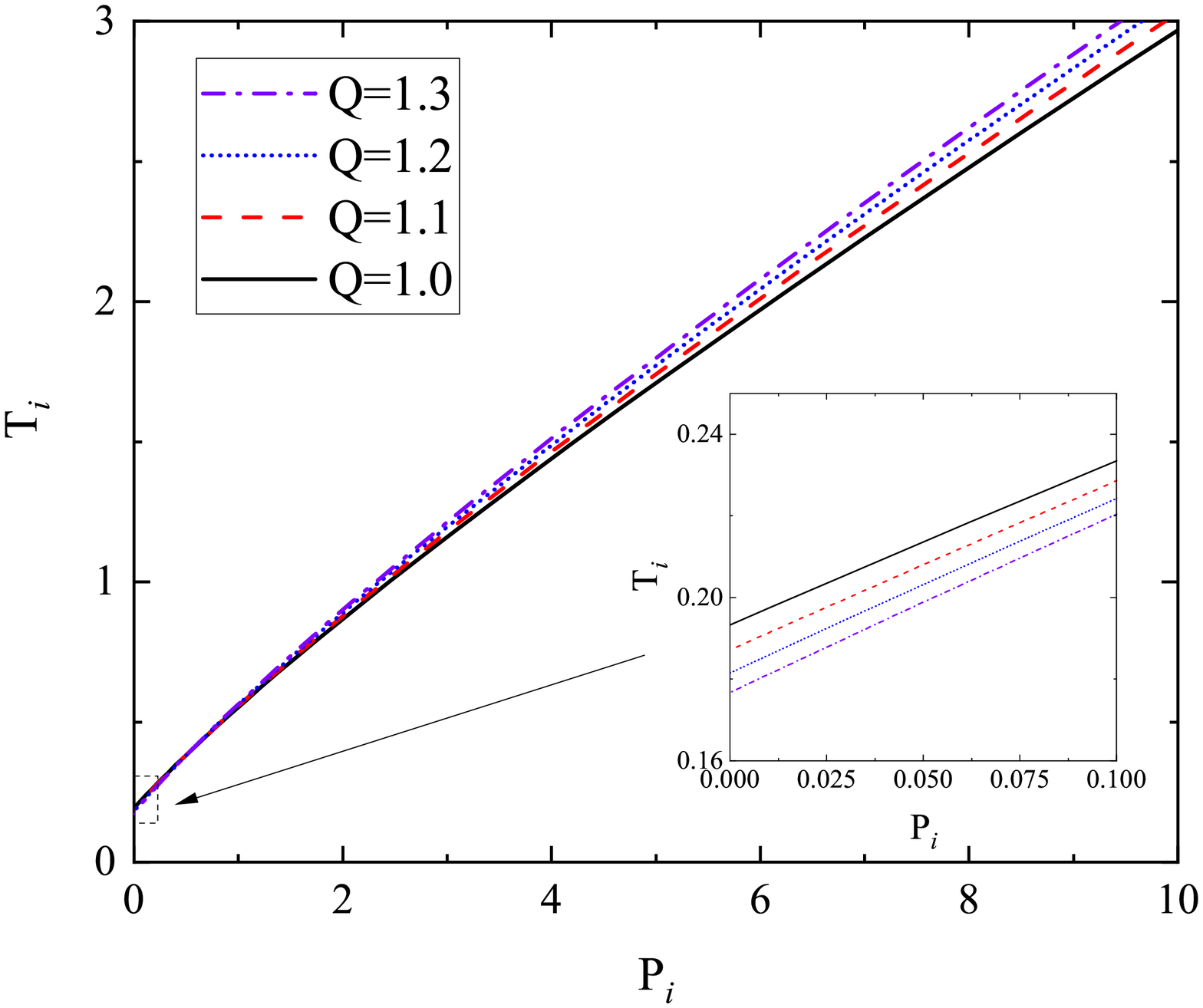}
\label{fig2-c}
\end{minipage}
}
\subfigure[$D=4,s=1.2$]{
\begin{minipage}[b]{0.31\textwidth}
\includegraphics[width=1.25\textwidth]{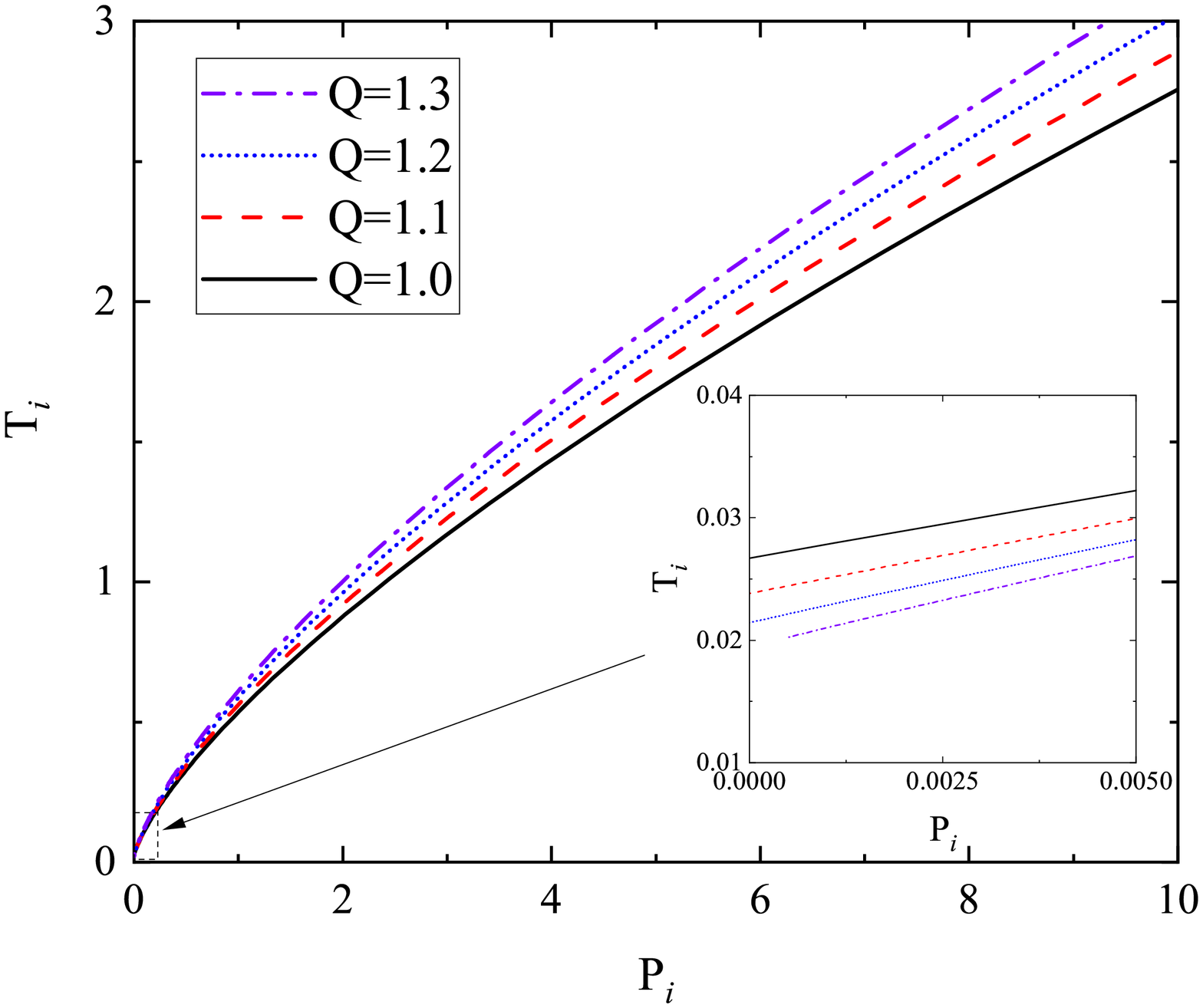}
\label{fig2-d}
\end{minipage}
}
\subfigure[$D=5,s=1.2$]{
\begin{minipage}[b]{0.31\textwidth}
\includegraphics[width=1.25\textwidth]{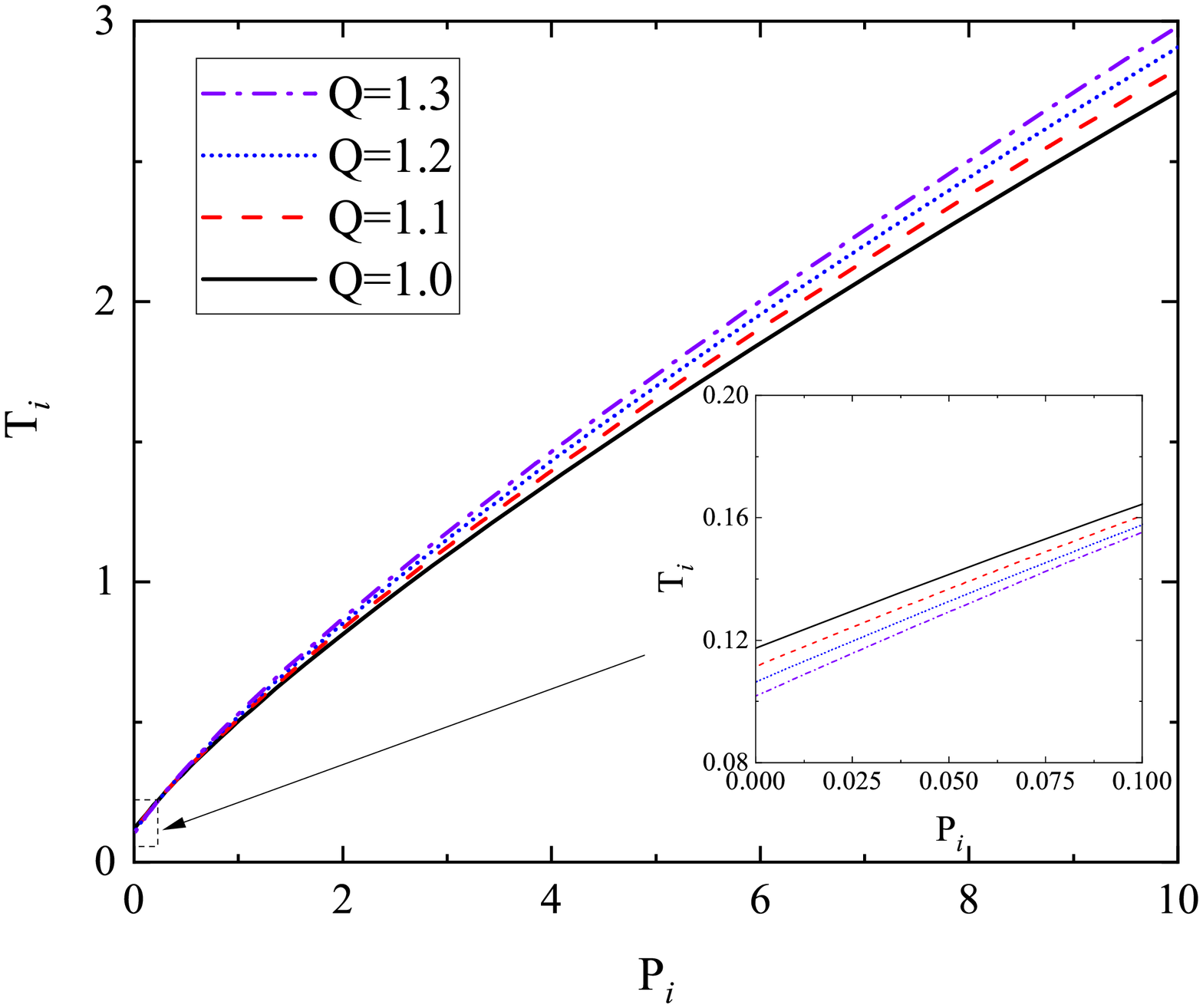}
\label{fig2-e}
\end{minipage}
}
\subfigure[$D=6,s=1.2$]{
\begin{minipage}[b]{0.31\textwidth}
\includegraphics[width=1.25\textwidth]{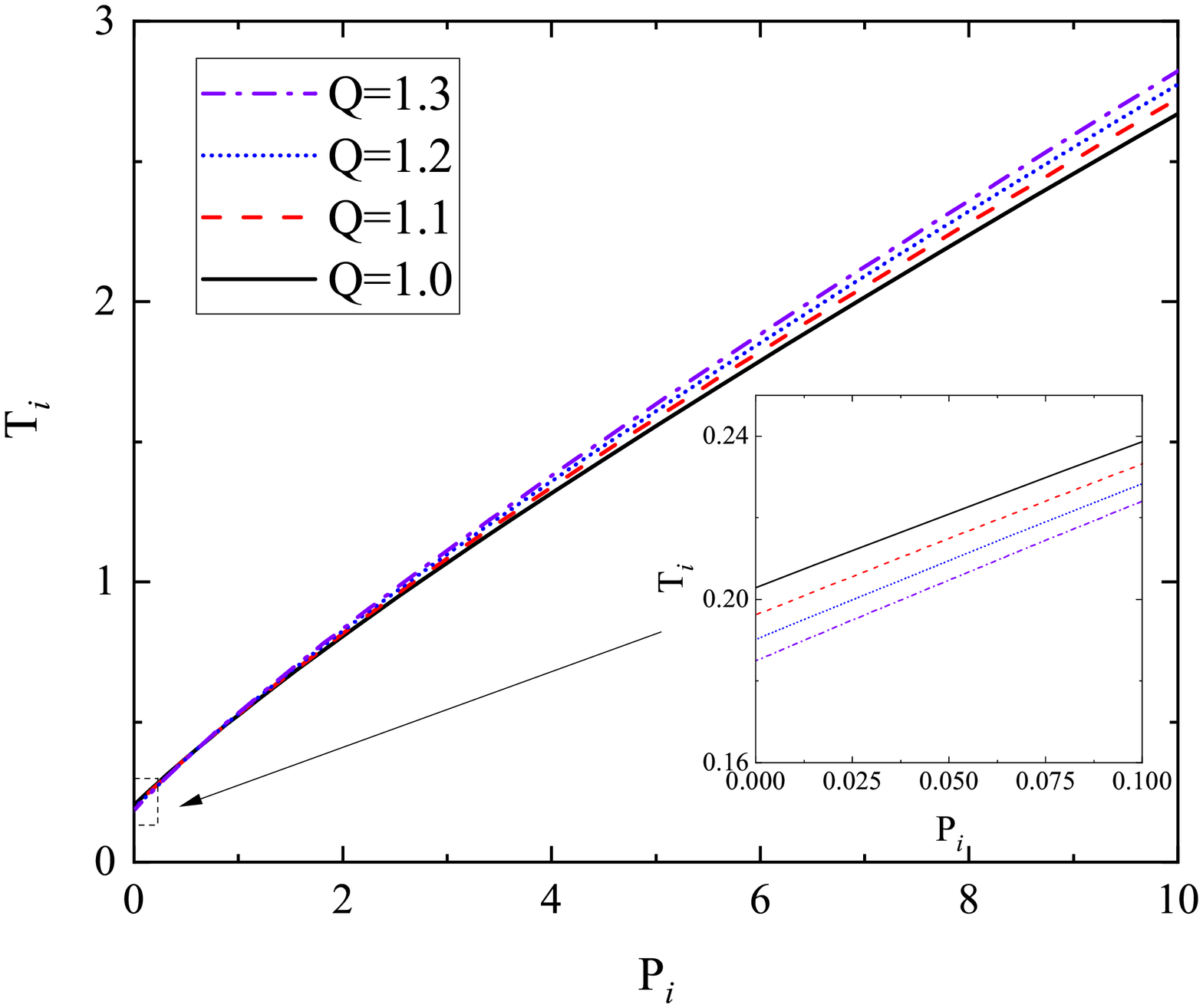}
\label{fig2-f}
\end{minipage}
}
\subfigure[$D=4,s=1.3$]{
\begin{minipage}[b]{0.31\textwidth}
\includegraphics[width=1.25\textwidth]{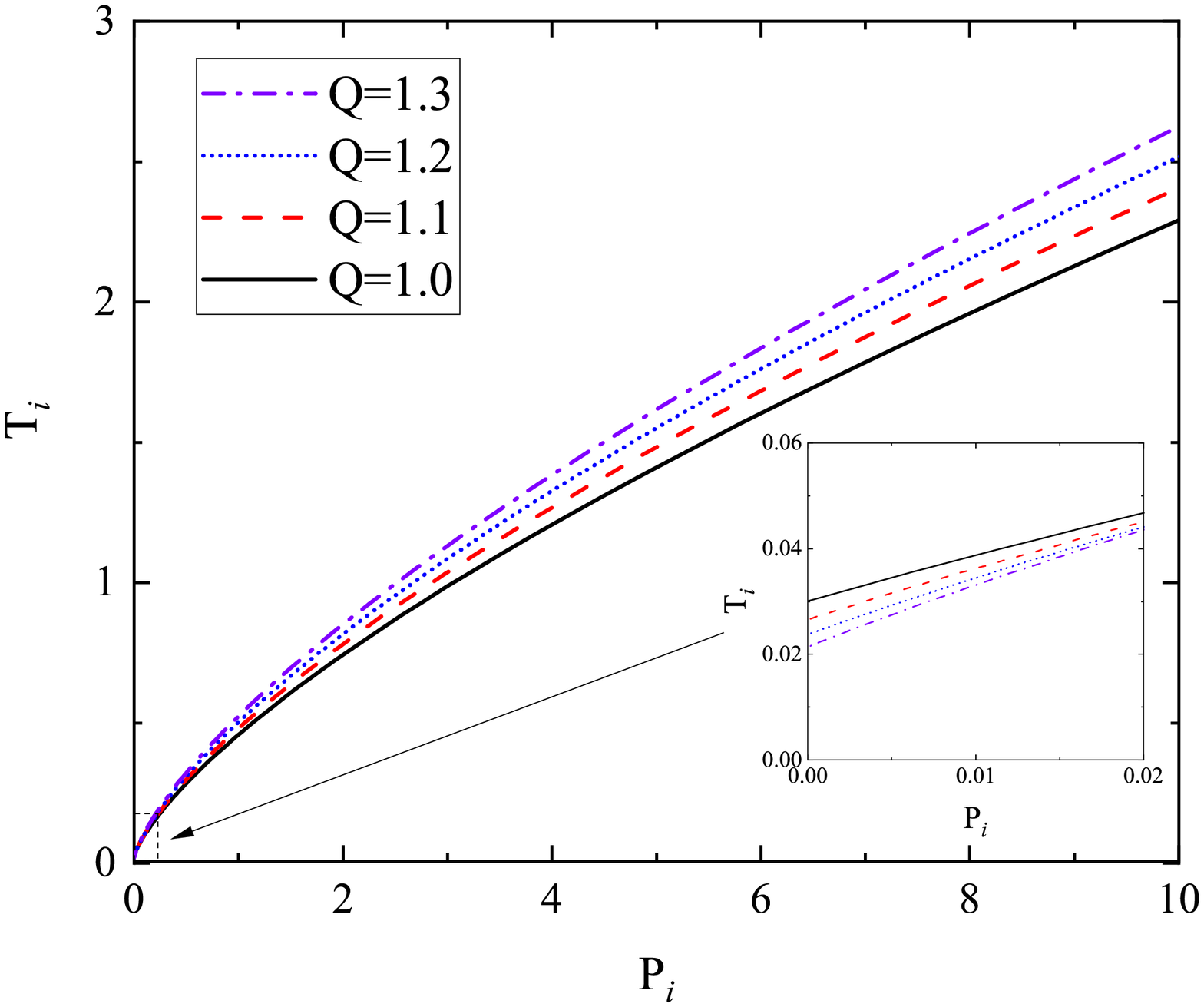}
\label{fig2-g}
\end{minipage}
}
\subfigure[$D=5,s=1.3$]{
\begin{minipage}[b]{0.31\textwidth}
\includegraphics[width=1.25\textwidth]{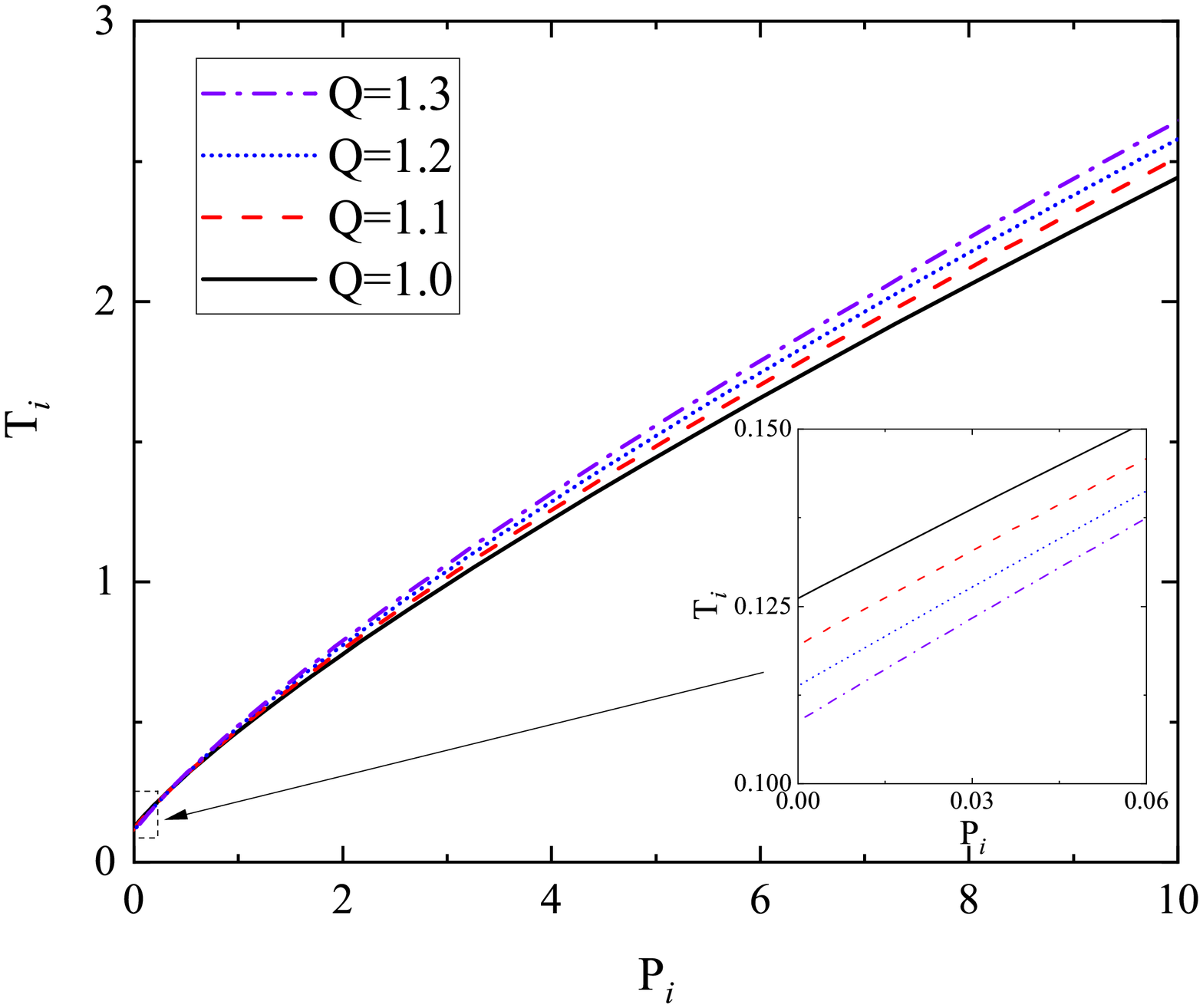}
\label{fig2-h}
\end{minipage}
}
\subfigure[$D=6,s=1.3$]{
\begin{minipage}[b]{0.31\textwidth}
\includegraphics[width=1.25\textwidth]{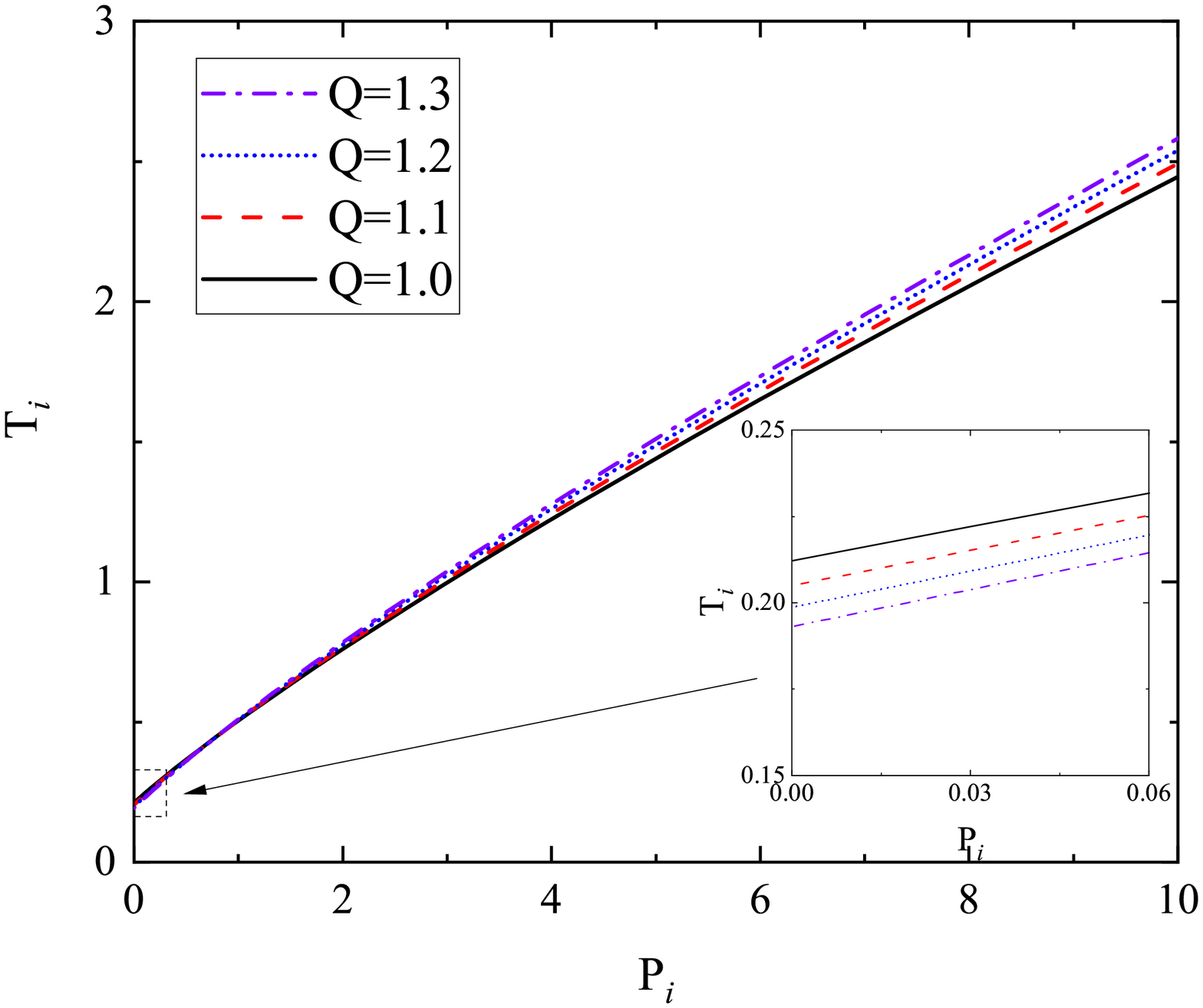}
\label{fig2-i}
\end{minipage}
}
\caption{The inversion curves for various combinations of $D$, $Q$  and  $s$.}
\label{fig2}
\end{figure*}

Next, by setting  ${\mu _{{\rm{BH}}}}=0$, one can the inversion pressure $P_i$, and then, substituting $P_i$ into Eq.~(\ref{eq9}), the  parameter equation of the inversion curve can be expressed follows \cite{cha35}
\begin{equation}
 \label{eq18}
\begin{cases}
{T_i} =\frac{1}{{2\pi \left( {D - 2} \right){r_ + }}}\left[ {3 - D + sr_ + ^{\frac{{2 + 2s\left( {D - 4} \right)}}{{1 - 2s}}}\Theta } \right],\\
{P_i} = \frac{1}{{16\pi r_ + ^2}}\left[ {\left( {3 - D} \right)D + r_ + ^{\frac{{2 + 2s\left( {D - 4} \right)}}{{1 - 2s}}}\left( {4s - 1} \right)\Theta } \right].
\end{cases}
\end{equation}
For further investigation of the inversion curve of the higher dimensional nonlinearly charged AdS black hole with PMI source, we plot Fig.~\ref{fig2} in the $T- P$ plane.

Fig.~\ref{fig2} illustrates the inversion temperature $T_i$ associated with inversion pressure $P_i$ for various combinations of $D$, $Q$  and  $s$. It is worth noting that  ${T_i}$ increases monotonously with ${P_i}$, this leads to each curve in the diagrams only exist one minimum value of inversion temperature  $T_i^{\min }$, and the cooling region and the heating region are located above and below these curves, respectively, which are dif\/ferent from that of Van der Waals f\/luids. By comparing the subgraphs of each row (e.g., Fig.~\ref{fig2-a}-Fig.~\ref{fig2-c}) and those of each column (e.g., Fig.~\ref{fig2-a}, Fig.~\ref{fig2-d} and Fig.~\ref{fig2-g}), respectively, it can be also found that, for high pressure, the $T_i$ decreases with the dimensionality $D$ and the nonlinearity parameter $s$, whereas it increases with charge $Q$. However, as seen from the small box in each subgraph, the behaviors of inversion curves at low pressure are reverse to those at high pressures. In other word,  $D$ and$s$ can enhance the curves of inversion temperature, while those curves decrease with $Q$.

Now, by demanding ${P_i} $ of Eq.~(\ref{eq18}) equals to zero, the roots are given by
\begin{align}
 \label{eq19}
{r_{\min }}  = {\left[ {\frac{{\left( {4s - 1} \right)\Theta }}{{D\left( {D - 3} \right)}}} \right]^{\frac{{2s - 1}}{{1 + \left( {D - 4} \right)s}}}},   {r'_{\min }} = - {\left[ {\frac{{\left( {4s - 1} \right)\Theta }}{{D\left( {D - 3} \right)}}} \right]^{\frac{{2s - 1}}{{1 + \left( {D - 4} \right)s}}}}.
\end{align}
It is worth noting  ${r'_{\min }}$ should be neglected  since it always negative. Substituting ${r_{\min }}$  into ${T_i}$  of Eq.~(\ref{eq18}), the minimum of inversion temperature can be obtained as
\begin{align}
 \label{eq20}
T_i^{\min }= & - \frac{{\left( {D - 3} \right)}}{{2\pi \left( {D - 2} \right)}}{\left[ {\frac{{\left( {4s - 1} \right)\Theta }}{{D\left( {D - 3} \right)}}} \right]^{\frac{{1 - 2s}} {{\left[ {1 + s\left( {D - 4} \right)} \right]}}}} 
  \nonumber \\
& + \frac{{s\Theta }}{{2\pi \left( {D - 2} \right)}}{\left[ {\frac{{\left( {4s - 1} \right)\Theta }}{{D\left( {D - 3} \right)}}} \right]^{ - \frac{{1 + 2s\left( {D - 3} \right)}}{{\left[ {1 + s\left( {D - 4} \right)} \right]}}}}.
\end{align}
Utilizing Eq.~(\ref{eq14}), the ratio between the minimum of inversion temperature and the critical temperature is
\begin{align}
 \label{eq21}
{\eta _{{\rm{BH}}}}  = & \frac{{T_i^{\min }}}{{{T_{cr}}}} =  \frac{{{{\left( {Ds} \right)}^{\frac{{2s - 1}}{{2 + 2s\left( {D - 4} \right)}}}}}}{{2\left( {D - 2} \right)}}{\left( {4s - 1} \right)^{ - \frac{{1 + 2s\left( {D - 3} \right)}}{{2 + 2s\left( {D - 4} \right)}}}} 
  \nonumber \\
 & \times {\left( {2s - 1} \right)^{\frac{{1 - 2s}}{{2 + 2s\left( {D - 4} \right)}}}} {\left[ {1 + 2s\left( {D - 3} \right)} \right]^{\frac{{1 + 2s\left( {D - 3} \right)}}{{2 + 2s\left( {D - 4} \right)}}}}.
\end{align}

In previous work, it is found that the ratio of R-N AdS black hole equals ${1 \mathord{\left/ {\vphantom {1 2}} \right. \kern-\nulldelimiterspace} 2}$, which indicates that the changing trend of ${T_i^{\min }}$ and ${{{T_{cr}}}}$ does not af\/fect by the  $Q$ \cite{cha16}. However, this f\/inding is not universal. Specif\/ic to our work, one can see that the ratio ${\eta _{{\rm{BH}}}}$  depends on the dimensionality  $D$ and the nonlinearity parameter $s$. When $s=1$, the ratio recovers the characteristic of the higher R-N AdS black hole system \cite{cha18}. For the sake of simplicity, we present the ratio with $s=1,3,5,7$  and  $D=4,5,6,7,\infty$ in Table.~\ref{tab1}.

\begin{table*}[htbp]
\centering
\caption{\label{tab1} The ratio ${\eta _{\rm {BH}}}$  for various nonlinearity parameter $s$ and dimensions $D$.}
\begin{tabular}{c  c  c  c}
\hline
Nonlinearity parameter                  &       Dimensions                               &      ${\eta _{\rm {BH}}}$ \\
\hline
\multirow{5}{*}{$s=1$}                 & $D=4$                                                &       0.500000 \\
                                                                  &$ D=5 $                                               &       0.471957 \\
                                                                  &$ D=6 $                                               &       0.452802\\
                                                                  &$ D=7 $                                               &       0.438933\\
                                                                  &$ D\rightarrow\infty $                &       0.333333\\
\multirow{5}{*}{$s=3$}                 & $D=4$                                                &       0.458603\\
                                                                  &$ D=5 $                                               &       0.434455 \\
                                                                  &$ D=6 $                                               &       0.414689\\
                                                                  &$ D=7 $                                               &       0.399481\\
                                                                  &$ D\rightarrow\infty $                &       0.272727\\
\multirow{5}{*}{$s=5$}                 & $D=4$                                                &       0.449778\\
                                                                  &$ D=5 $                                               &       0.427255 \\
                                                                  &$ D=6 $                                               &       0.407755\\
                                                                  &$ D=7 $                                               &       0.392523\\
                                                                  &$ D\rightarrow\infty $                &       0.263158\\
\multirow{5}{*}{$s=7$}                 & $D=4$                                                &       0.445956\\
                                                                  &$ D=5 $                                               &       0.424204 \\
                                                                  &$ D=6 $                                               &       0.404849\\
                                                                  &$ D=7 $                                               &       0.389624\\
                                                                  &$ D\rightarrow\infty $                &       0.259259\\
\hline
\end{tabular}
\end{table*}

\begin{figure}[htbp]
\centering 
\includegraphics[width=.45\textwidth,origin=c,angle=0]{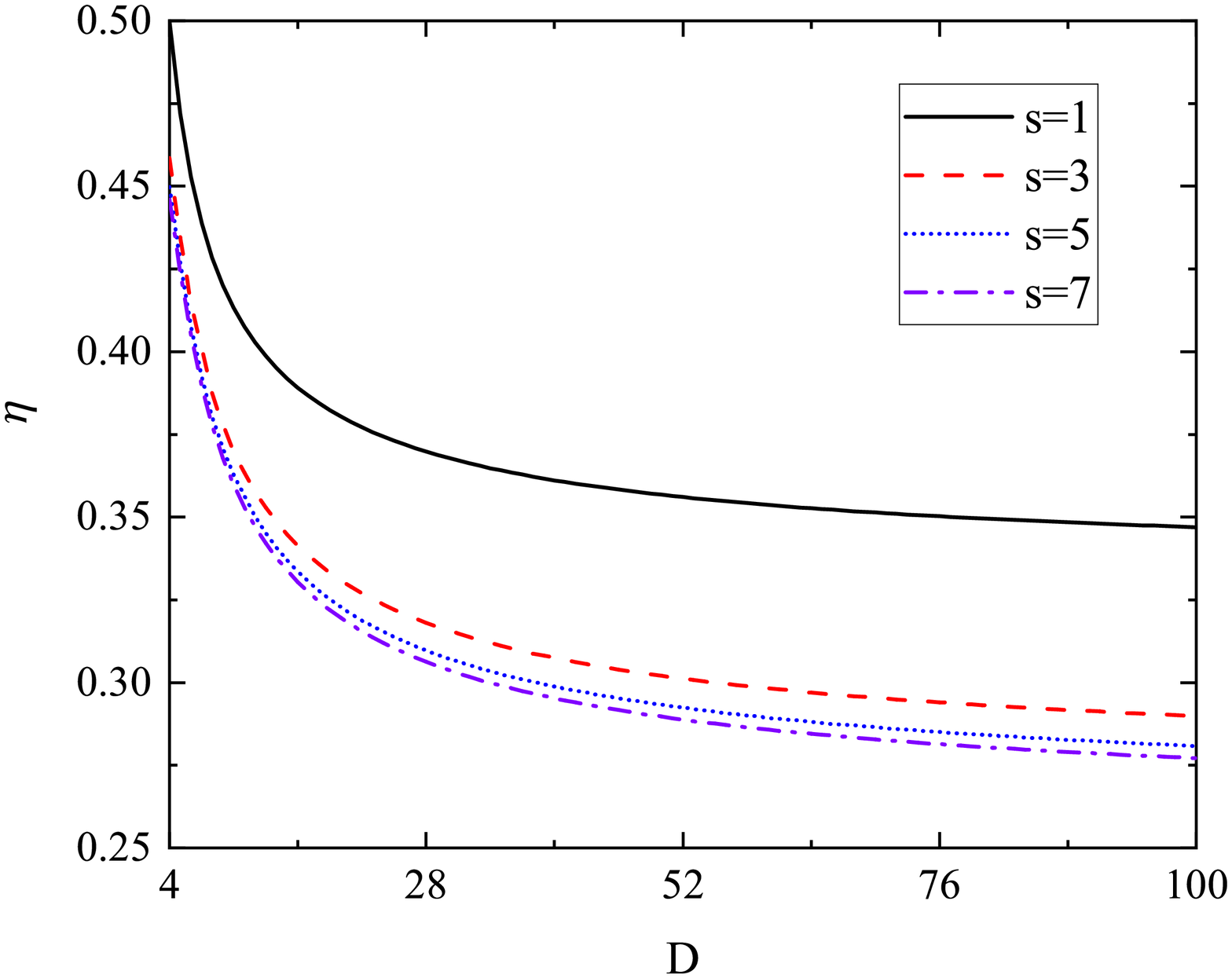}
\caption{\label{fig3} Relationship between the ratio  ${\eta _{\rm {BH}}}$ and the dimension $D$   for dif\/ferent  $s$.}
\end{figure}

In Table.~\ref{tab1},   the ratio between the minimum of inversion temperature and the critical temperature decrease with  $s$ and $D$, which means that the denominator  ${T_{cr}}$ is grows faster than the numerator  ${T_{\min }}$. When $D \rightarrow \infty$, the curve of  ${\eta _{\rm {BH}}}$ approaches a constant. Interestingly, according to Fig.~\ref{fig3}, one can f\/ind that the two adjacent curves get closer and closer with the increase of  $s$, which never appear in the previous works.

\begin{figure*}[htbp]
\centering
\subfigure[$D=4,s=1.1,Q=1$]{
\begin{minipage}[b]{0.31\textwidth}
\includegraphics[width=1.25\textwidth]{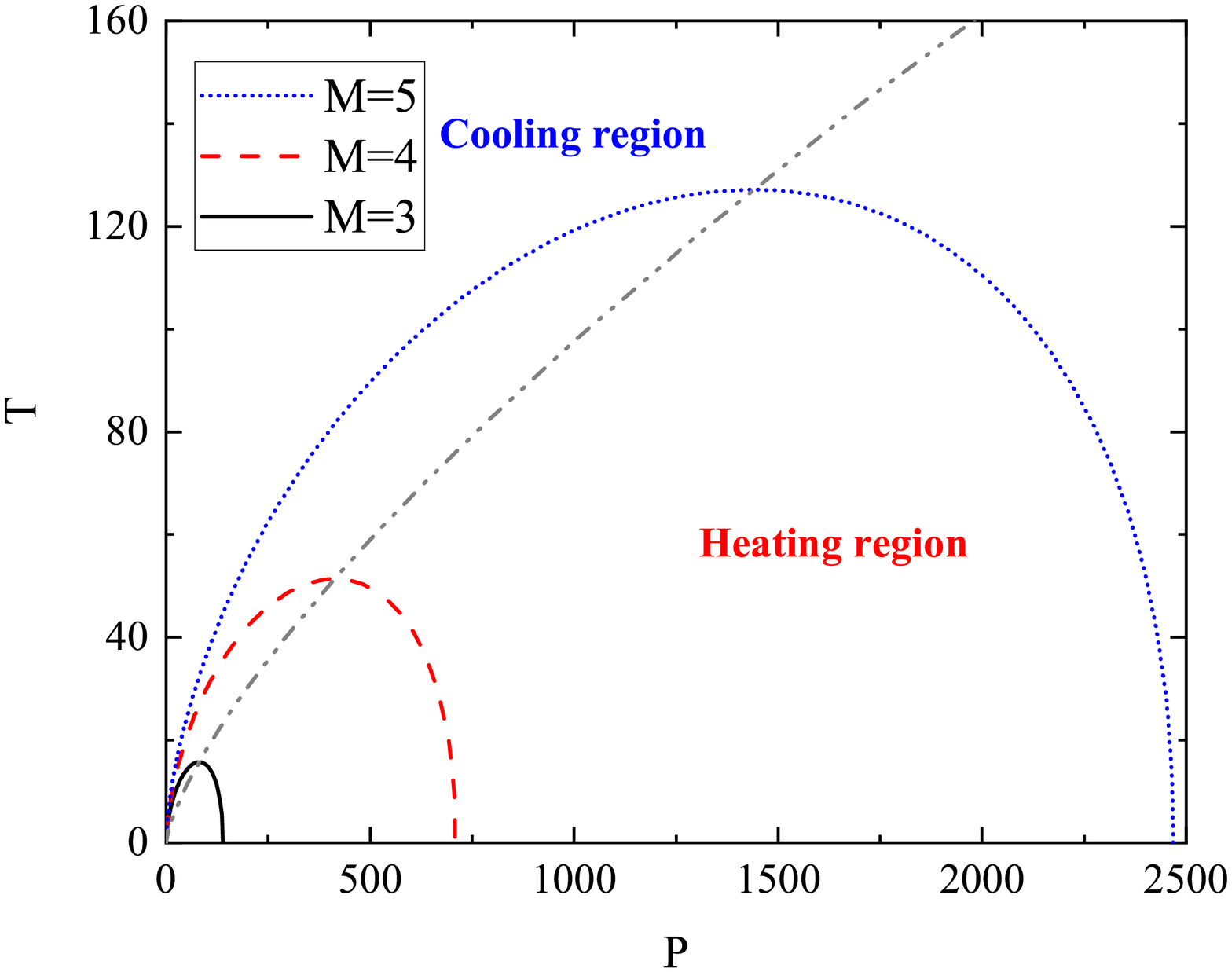}
\label{fig4-a}
\end{minipage}
}
\subfigure[$D=5,s=1.1,Q=1$]{
\begin{minipage}[b]{0.31\textwidth}
\includegraphics[width=1.25\textwidth]{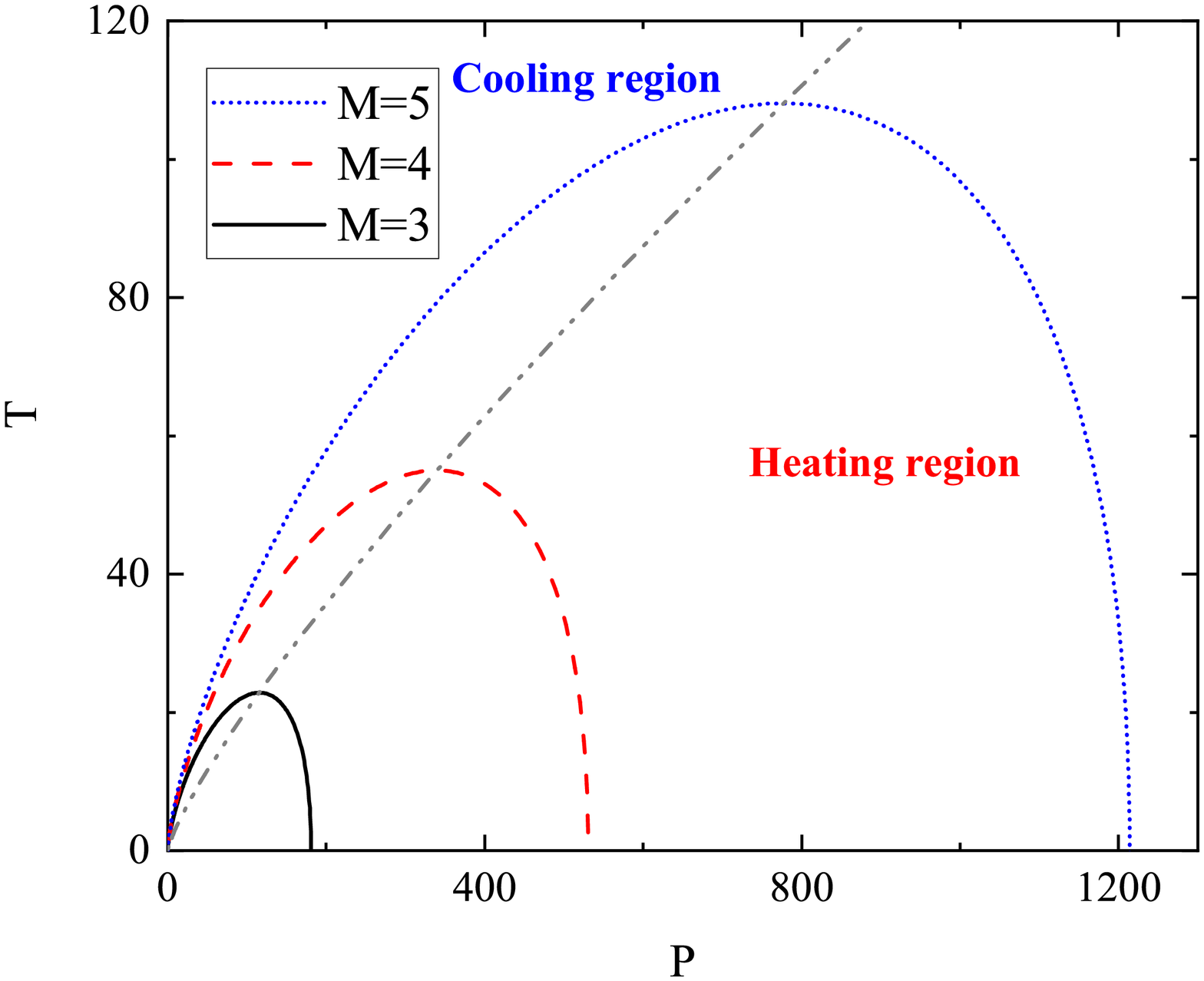}
\label{fig4-b}
\end{minipage}
}
\subfigure[$D=6,s=1.1,Q=1$]{
\begin{minipage}[b]{0.31\textwidth}
\includegraphics[width=1.25\textwidth]{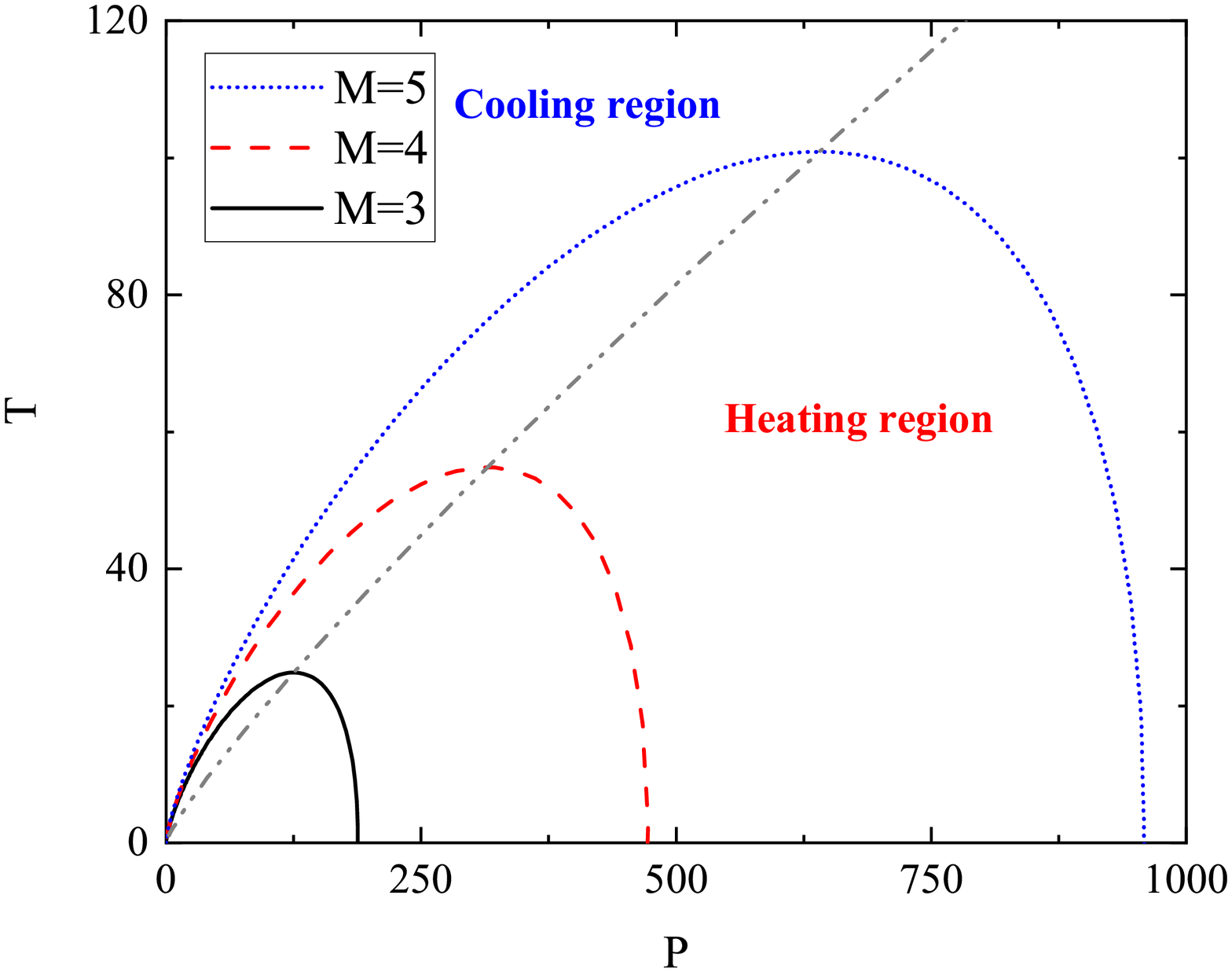}
\label{fig4-c}
\end{minipage}
}
\subfigure[$D=4,s=1.2,Q=1$]{
\begin{minipage}[b]{0.31\textwidth}
\includegraphics[width=1.25\textwidth]{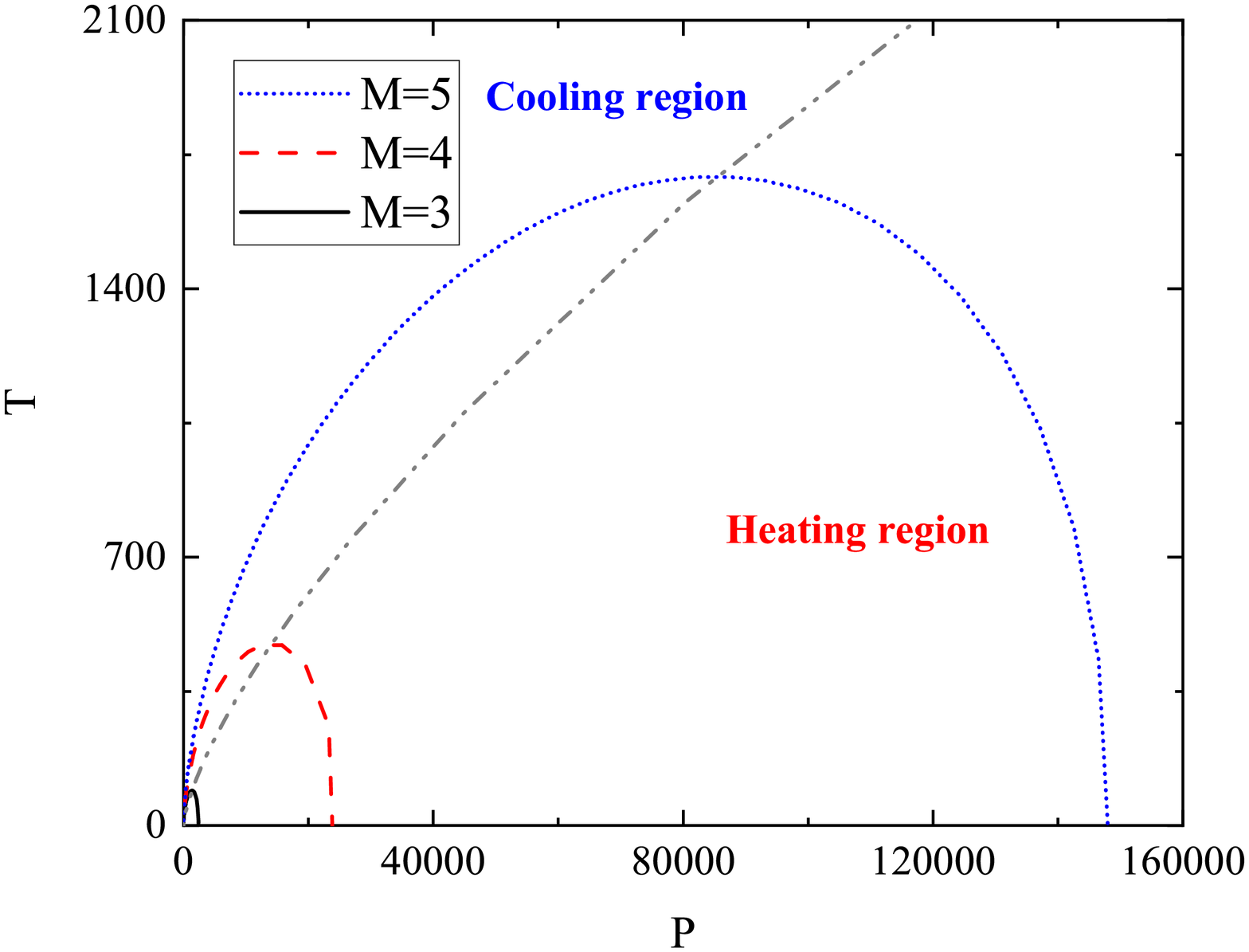}
\label{fig4-d}
\end{minipage}
}
\subfigure[$D=5,s=1.2,Q=1$]{
\begin{minipage}[b]{0.31\textwidth}
\includegraphics[width=1.25\textwidth]{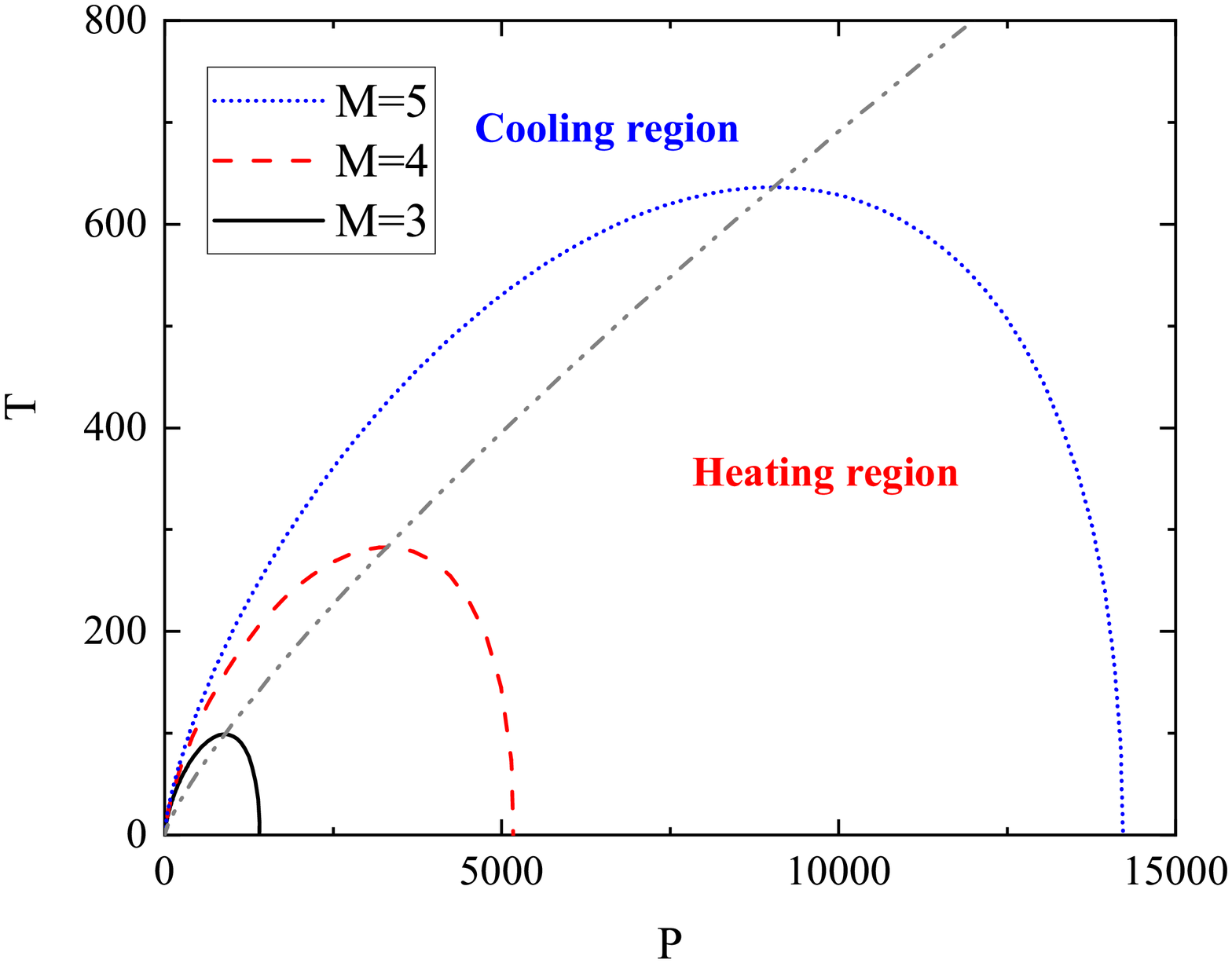}
\label{fig4-e}
\end{minipage}
}
\subfigure[$D=6,s=1.2,Q=1$]{
\begin{minipage}[b]{0.31\textwidth}
\includegraphics[width=1.25\textwidth]{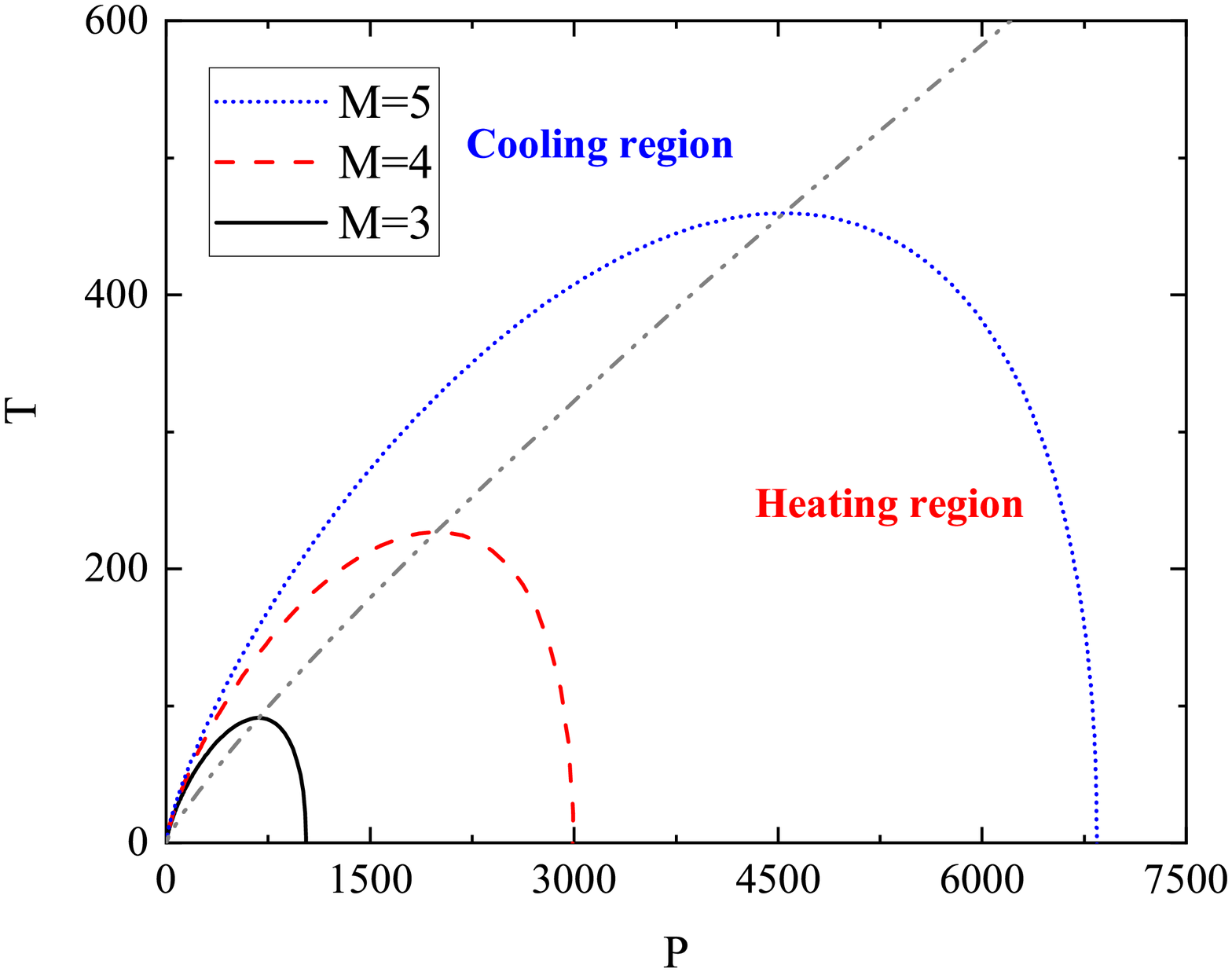}
\label{fig4-f}
\end{minipage}
}
\subfigure[$D=4,s=1.2,Q=1.1$]{
\begin{minipage}[b]{0.31\textwidth}
\includegraphics[width=1.25\textwidth]{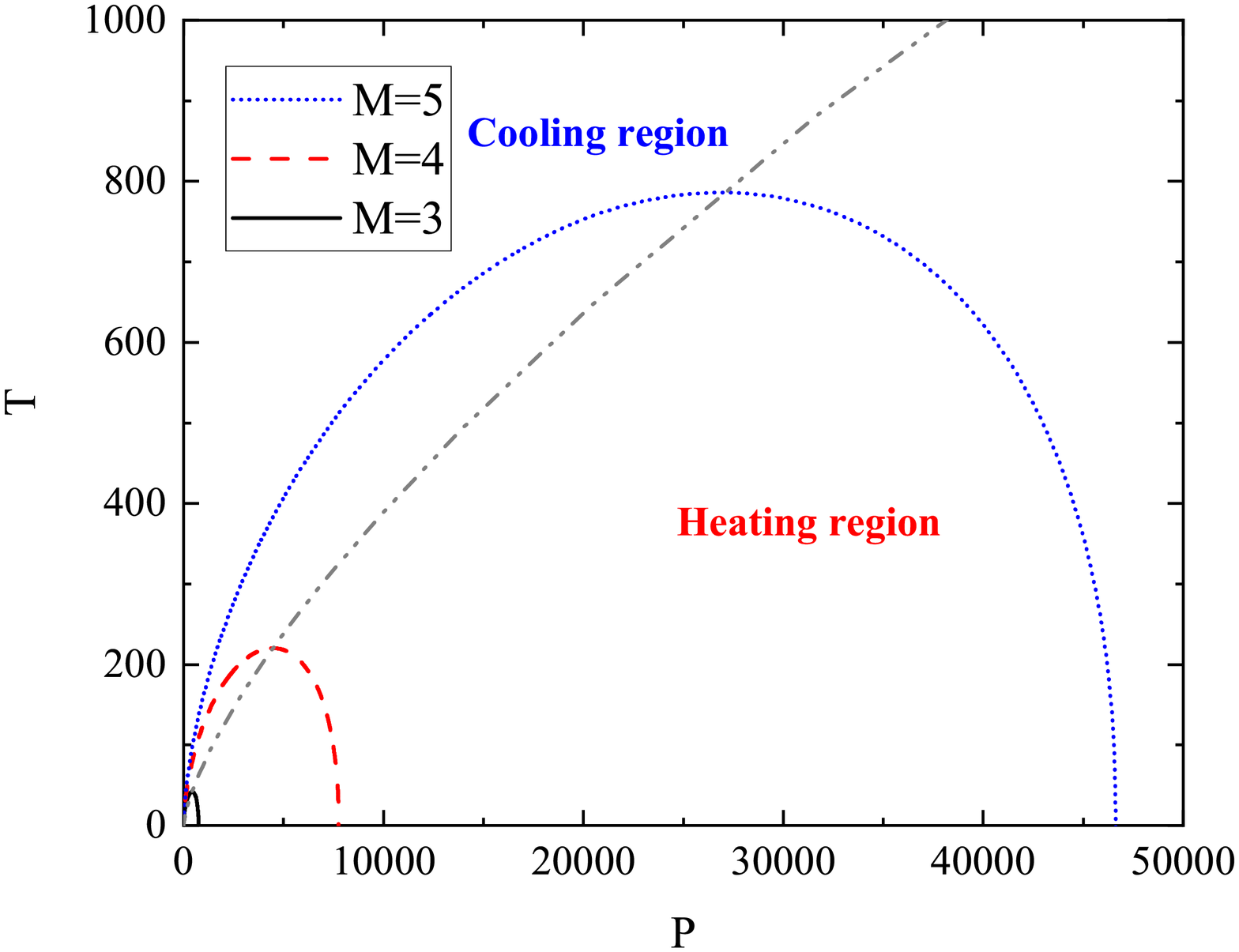}
\label{fig4-g}
\end{minipage}
}
\subfigure[$D=5,s=1.2,Q=1.1$]{
\begin{minipage}[b]{0.31\textwidth}
\includegraphics[width=1.25\textwidth]{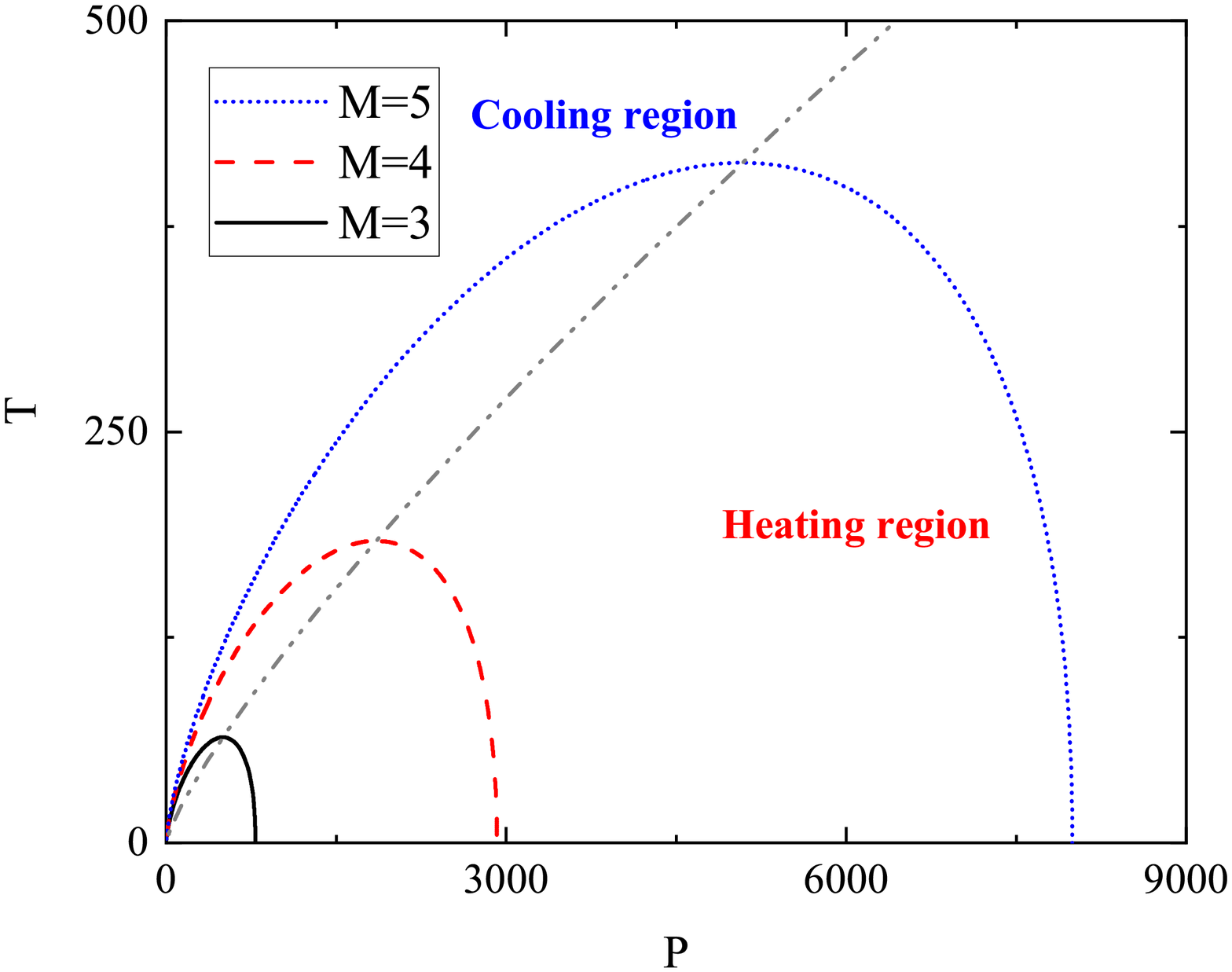}
\label{fig4-h}
\end{minipage}
}
\subfigure[$D=6,s=1.2,Q=1.1$]{
\begin{minipage}[b]{0.31\textwidth}
\includegraphics[width=1.25\textwidth]{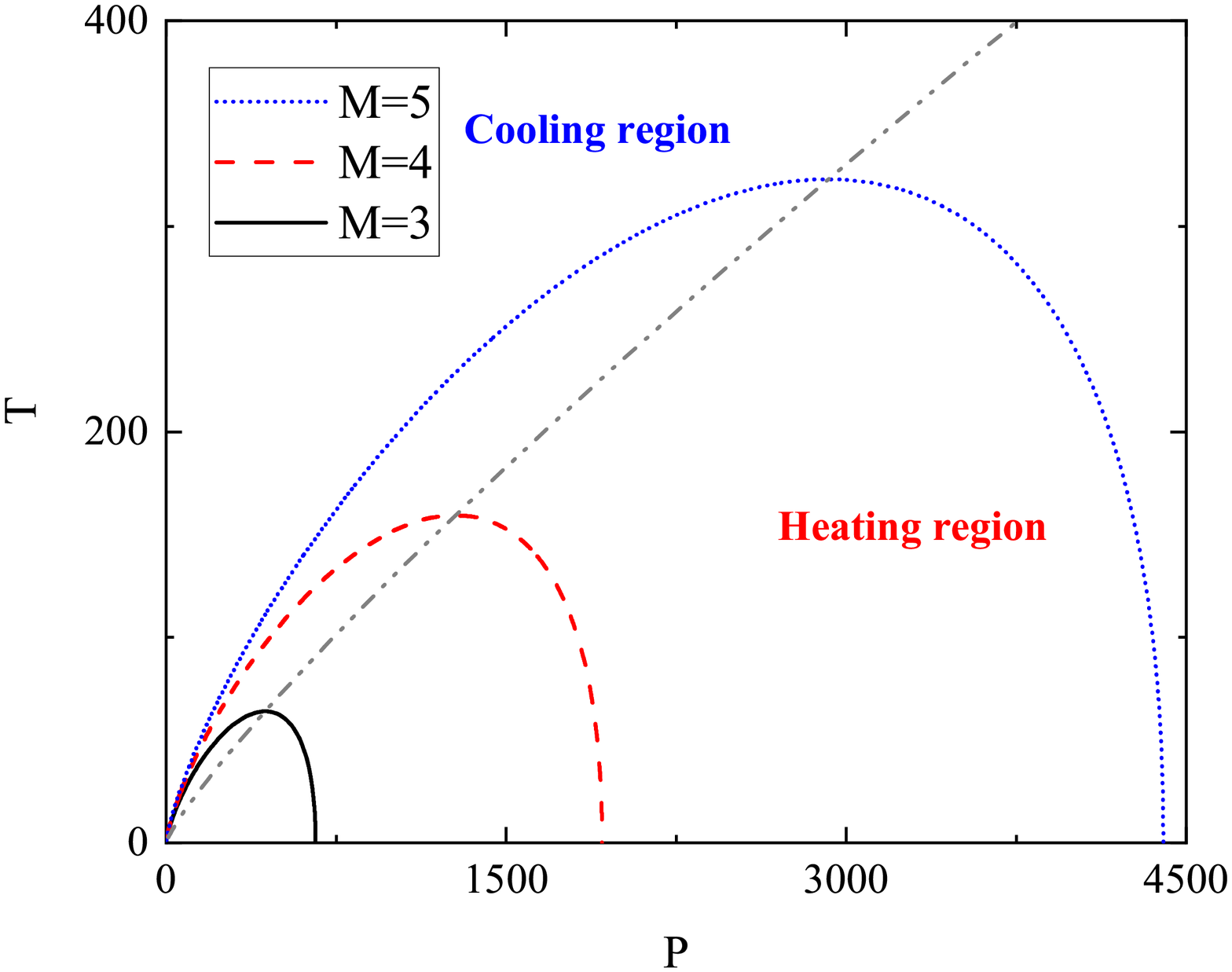}
\label{fig4-i}
\end{minipage}
}
\caption{The isenthalpic curves for various combinations of $D$, $s$ and $Q$.}
\label{fig4}
\end{figure*}

Finally, considering the Joule-Thomson expansion is an isenthalpic process, it is interesting to investigate the isenthalpic curves in $T-P$ plane. According to  Eq.~(\ref{eq8}) and Eq.~(\ref{eq13}), the equation of the isenthalpic curves is given by

\begin{equation}
 \label{eq20}
 {}
\begin{cases}
T =   - \frac{1}{{2\pi {r_ + }}} + \frac{{4M\left( {D - 1} \right)}}{{\left( {D - 2} \right){\omega _{D - 2}}r_ + ^{D - 2}}} + \frac{{sr_ + ^{\frac{{1 + 2\left( {D - 3} \right)s}}{{1 - 2s}}}{{\left( {2s - 1} \right)}^s}}}{{2\pi \left( {1 - D + 2s} \right)}}\Theta ,\\
P =  - \frac{{\left( {D - 1} \right)\left( {D - 2} \right)}}{{16\pi r_ + ^2}} + \frac{{\left( {D - 1} \right)M}}{{{\omega _{D - 2}}r_ + ^{D - 1}}} - \frac{{\left( {D - 1} \right){{\left( {1 - 2s} \right)}^2}r_ + ^{ - \frac{{2s(D - 2)}}{{2s - 1}}}}}{{16\pi \left( {D - 2s - 1} \right)}}\Theta .
\end{cases}
\end{equation}
By using Eq.~(\ref{eq21}) and considering the ADM mass of the black hole is equals to its enthalpy in the extended phase space, that is  $H=M$, the isenthalpic curves for various combinations of $D$, $s$, and $Q$ are represented in Fig.~\ref{fig4}.

In Fig.~\ref{fig4}, each graphics has three isenthalpic curves for dif\/ferent mass, the black solid curve, red dashed curve, and blue dotted curve are $M=3$, $M=4$, and $M=5$, respectively. The gray dot-dash curve represents the inversion curve, which is consistent with that in Fig.~\ref{fig2}. The inversion curves  intersect the maximum point of the isenthalpic curves, it naturally leads to the left-hand side of the isenthalpic curve has a positive slope, while the slope of the isenthalpic curve becomes negative at the right-hand side. Therefore, in the throttling process, the inversion curve is regarded as the dividing line between the heating region and the cooling region. Meanwhile, by analyzing and comparing, it is easy to see that increase the mass $M$ and $s$, or reduce the charge $Q$ and $D$ can enhance the isenthalpic curve. Furthermore, comparing Fig.~\ref{fig4-a} with Fig.~\ref{fig4-e},  one can f\/ind that the isenthalpic curve still expands rightward when increasing the $D$, $Q$, and $s$ at the same time. This indicates that the ef\/fect of $s$ on the isenthalpic curve is much greater than other parameters.

 \section{Conclusion and Discussion}
\label{Dis}
In this paper, by considering cosmological constant as the pressure, we investigated Joule-Thomson expansion of the higher dimensional nonlinearly charged AdS black hole with PMI source. Firstly, according to the thermodynamic quantities of the higher dimensional nonlinearly charged AdS black hole with PMI source, we calculated the Joule-Thomson coef\/f\/icient  ${\mu _{{\rm{BH}}}}$.  The results showed that the ${\mu _{{\rm{BH}}}}$  is related to the dimensionality $D$, charge $Q$ and nonlinearity parameter $s$. Meanwhile, it has a zero point and a divergent point, which are coincide the inversion temperature  ${T_i}$ and the zero point of Hawking temperature, respectively. The curve of the Joule-Thomson coef\/f\/icient moves to the right as the spacetimes $D$ or $Q$ increases, while it moves to the left as  $s$ increases. Secondly, we analyzed inversion curve via Eq.~(\ref{eq18}) and Fig.~\ref{fig2}. It is found that  $T_i$ increases monotonously with  $P_i$ and leads to only one minimum value of inversion temperature $T_i^{\min }$ in the black hole system. At the high pressure area, the inversion curves  increase as the dimensionality and the nonlinearity parameter decrease, or the charge increase. However, at the low pressure area, $T_i$  increase as the $D$ and $s$ increase, or $Q$ decrease.  Next, we derived the expression of  $T_i^{\min }$, and calculate the ratio between the minimum inversion temperature and the critical temperature  ${\eta _{BH}}$. Both $T_i^{\min }$  and  ${T_{cr}}$ all contain the charge, whereas the ratio ${\eta _{BH}}$  between them has nothing to do with  $Q$. For  $s=1$,  ${\eta _{BH}}$ recovers the characteristic of the higher R-N AdS black hole system. If  $s>1$, it becomes smaller and smaller as  $D$ increases, which means that the denominator  $T_{cr}$ grows faster than the numerator $T_{\min }$.  According to Fig.~\ref{fig3}, it is obvious that the two adjacent curves get closer and closer with the increase of  $s$, which never appear in the previous works.  Finally, according to Eq.~(\ref{eq21}) and considering $M=H$  in the extended phase space, we plot the isenthalpic curves for various combinations of $D$, $M$, $Q$ and $s$ in Fig.~\ref{fig4}. It is easy to see that increase the mass $M$ and $s$, or reduce the charge $Q$ and $D$ can enhance the isenthalpic curve, and the ef\/fect of $s$ on the isenthalpic curve is much greater than other parameters.



\end{document}